\documentclass[reqno,12pt]{article}
\usepackage{ifpdf}
\usepackage{ae}
\usepackage[T1]{fontenc}
\usepackage[ansinew]{inputenc}
\usepackage{mathrsfs}
\usepackage{amsmath}
\usepackage{amssymb}
\usepackage{graphicx}
\usepackage{color}
\definecolor{darkblue}{cmyk}{0.9,0.9,0,0}
\definecolor{darkgreen}{rgb}{0,0.55,0}
\usepackage[colorlinks=true,linkcolor=darkblue,citecolor=darkblue,urlcolor=darkblue]{hyperref}
\usepackage{epsfig}

\newcommand{\beq}{\begin{equation}}
\newcommand{\eeq}{\end{equation}}
\newcommand{\beqq}{\begin{equation*}}
\newcommand{\eeqq}{\end{equation*}}
\newcommand\beqa{\begin{eqnarray}}
\newcommand\eeqa{\end{eqnarray}}
\newcommand\beqaa{\begin{eqnarray*}}
\newcommand\eeqaa{\end{eqnarray*}}
\newcommand\bea{\begin{array}}
\newcommand\eea{\end{array}}

\def\XXint#1#2#3{{\setbox0=\hbox{$#1{#2#3}{\int}$}
\vcenter{\hbox{$#2#3$}}\kern-.5\wd0}}

\newcommand{\nn}{\nonumber}
\newcommand{\com}[1]{(*{\textbf{#1}}*)}

\newcommand{\neqa}{\nonumber\end{eqnarray}}
\newcommand{\la}[1]{\label{#1}}

\newcommand{\eq}[1]{eq.(\ref{#1})}

\newcommand{\B}{\mathcal{B}}

\newcommand{\Tr}{{\rm Tr}}

\renewcommand{\d}{\partial}

\newcommand{\<}{{\langle}}
\renewcommand{\>}{{\rangle}}

\newcommand{\re}{\relax{\rm I\kern-.18em R}}

\newcommand{\caB}{{\mathscr B}}

\newcommand{\caA}{{\mathscr A}}
\newcommand{\caC}{{\mathscr C}}

\renewcommand{\sp}{p\hspace{-.40em}/}

\def\su2{{SU(2)}}

\def\a{{\alpha}}

\def\[{\left[}
\def\]{\right]}

\def\a{\alpha}

\def\({\left(}
\def\){\right)}
\def\[{\left[}
\def\]{\right]}

\def\<{\langle}
\def\>{\rangle}

\def\i2{\frac{i}{2}}

\newcommand{\N}{\mathcal{N}}
\def\O{{\mathcal O}}

\def\spi{\relax{\rm \pi\kern-0.5em /}}
\def\sA{\relax{\rm A\kern-0.5em /}}
\def\sp{\relax{\rm p\kern-0.5em /}}
\def\sd{\relax{\rm \d\kern-0.5em /}}
\def\sk{\relax{\rm k\kern-0.5em /}}
\def\sn{\relax{\rm n\kern-0.5em /}}
\def\sl{\relax{\rm l\kern-0.5em /}}
\def\sP{\relax{\rm P\kern-0.7em /}}
\def\sBethe{\relax{\rm \Bethe\kern-0.5em /}}

\def\co{\text{co}}
\def\al{\text{al}}

\def\bps{\text{BPS}}

\usepackage{varioref}
\usepackage{makeidx}
\makeindex

\usepackage[english]{babel}
\begin{document}

\thispagestyle{empty}

\renewcommand{\thefootnote}{\fnsymbol{footnote}}
\setcounter{page}{1}
\setcounter{footnote}{0}
\setcounter{figure}{0}
\begin{center}
$$$$
{\Large\textbf{\mathversion{bold} Tailoring Three-Point Functions and Integrability  }\par}

\vspace{1.0cm}

\textrm{Jorge Escobedo$^{a,b}$, Nikolay Gromov$^c$, Amit Sever$^a$ and Pedro Vieira$^{a}$}
\\ \vspace{1.2cm}
\footnotesize{

\textit{$^{a}$ Perimeter Institute for Theoretical Physics\\ Waterloo,
Ontario N2L 2Y5, Canada}  \\
\texttt{jescob,amit.sever,pedrogvieira@gmail.com} \\
\vspace{4mm}
\textit{$^{b}$ Department of Physics and Astronomy \& Guelph-Waterloo Physics Institute,\\
University of Waterloo, Waterloo, Ontario N2L 3G1, Canada} \\
\vspace{4mm}
\textit{$^c$ King's College London, Department of Mathematics WC2R 2LS, UK \& \\
St.Petersburg INP, St.Petersburg, Russia} \\
\texttt{nikgromov@gmail.com}
\vspace{3mm}
}

\par\vspace{1.5cm}

\textbf{Abstract}\vspace{2mm}
\end{center}

\noindent

We use Integrability techniques to compute structure constants in $\mathcal{N}=4$ SYM to leading order.
 Three closed spin chains, which represent the single trace gauge-invariant operators in $\mathcal{N}=4$ SYM,
 are cut into six open chains which are then sewed back together into some nice pants, the three-point function.
 The algebraic and coordinate Bethe ansatz tools necessary for this task are reviewed.
Finally, we discuss the classical limit of our results, anticipating some predictions for quasi-classical string correlators in terms of algebraic curves.

\vspace*{\fill}

\setcounter{page}{1}
\renewcommand{\thefootnote}{\arabic{footnote}}
\setcounter{footnote}{0}

\newpage

 \def\nref#1{{(\ref{#1})}}

\tableofcontents

\section{Introduction}
Solving interacting conformal field theories in $4d$ with a large $N$ expansion will have a deep impact in our understanding of Nature.
$\mathcal{N}=4$ super Yang-Mills (SYM) seems to be, excitingly, the harmonic oscillator of gauge theories in four dimensions. Its full solution will dramatically improve our understanding of particle theories such as QCD.

The fundamental objects in conformal field theories are two- and three-point functions of local gauge-invariant operators. Knowing them we can in principle construct any higher-point function by gluing these building blocks together.
Two-point functions are greatly understood in $\mathcal{N}=4$ SYM, largely due to the existence of integrability \cite{integrability,MZ}. In this paper we shall focus on the study of planar three-point functions, or structure constants, using the underlying exactly solvable structures of these theories. We will illustrate our methods at weak coupling in $\mathcal{N}=4$ SYM.

For some interesting and inspirational works on three-point function in $\mathcal{N}=4$ SYM at weak coupling see \cite{Okuyama:2004bd,Roiban:2004va,Alday:2005nd}. In particular, \cite{Okuyama:2004bd} introduces the physical picture of cutting and gluing spin chains which we elaborate on below and \cite{Roiban:2004va} emphasizes the usefulness of the algebraic Bethe ansatz techniques for computing scalar products of quantum spin chains, which turns out to be very relevant for this problem.

In any conformal field theory (CFT), one can choose a basis of local operators such that their two-point functions are given by
\beq\la{2pf}
\<\O_i(x)\bar\O_j(0)\>=\mathcal{N}_i{\delta_{ij}\over|x|^{2\Delta_i}}\ ,
\eeq
where $\Delta_i$ are their conformal dimensions and $\mathcal{N}_i$ are normalization constants that may be set to one. The correlation function of three such local operators is restricted by conformal symmetry to be of the form
\beq\la{3pt}
\<\O_i (x_i) \O_j (x_j) \O_k (x_k) \> = \frac{\sqrt{\mathcal{N}_i\,\mathcal{N}_j\,\mathcal{N}_k}\ C_{ijk}}{|x_{ij}|^{\Delta_i+\Delta_j-\Delta_k} |x_{jk}|^{\Delta_j+\Delta_k-\Delta_i} |x_{ki}|^{\Delta_k+\Delta_i-\Delta_j}},
\eeq
where $C_{ijk}$ are the structure constants.
In what follows we will compute these structure constants in the planar limit of $\mathcal{N}=4$ SYM to leading order in $1/N_c$ and $\lambda$, i.e. the number of colors and the 't Hooft coupling. The first nonzero structure constants in the $1/N_c$ expansion scale as $1/N_c$. These arise in the three-point function of single trace operators. Therefore, our operators will all be of the form $\Tr(ABC\dots)$ where $A,B,C$ are  $\mathcal{N}=4$ fields. These are the most basic correlation functions building blocks.\footnote{One would expect that in order to Bootstrap the four-point function of single trace operators, the knowledge of the three-point functions between two single trace operators and one higher trace operator would also be required. Remarkably, this does not seem to be the case; the four-point function seems to be re-constructable from the three-point function of single trace operators alone \cite{mack}.}

The corresponding single trace structure constants have a perturbative expansion of the form
\beq\la{expansion}
N_c \,C_{ijk}=c_{ijk}^{(0)}+ \lambda c_{ijk}^{(1)}+ \lambda^2 c_{ijk}^{(2)}+\dots
\eeq
and we will only consider the first term in this expansion $c_{ijk}^{(0)}$. At $\lambda=0$ many single trace operators have the same dimension, which is roughly the number of fundamental fields in the operator. That huge degeneracy is lifted at one loop. Therefore, to correctly identify the coefficient $c_{ijk}^{(0)}$ in the expansion (\ref{expansion}), we have to use those linear combinations of single trace operators that have definite one-loop anomalous dimension $\Delta_i=\Delta_i^{(0)}+\lambda\gamma_i^{(1)}$.\footnote{This is nothing but the standard textbook degenerate perturbation theory in QM and needs to be taken into account. See e.g. \cite{Beisert:2002bb} and \cite{revieC} for a discussion of the importance of this point. }

\begin{figure}[t]
\centering
\def\svgwidth{16cm}

\ifpdf
    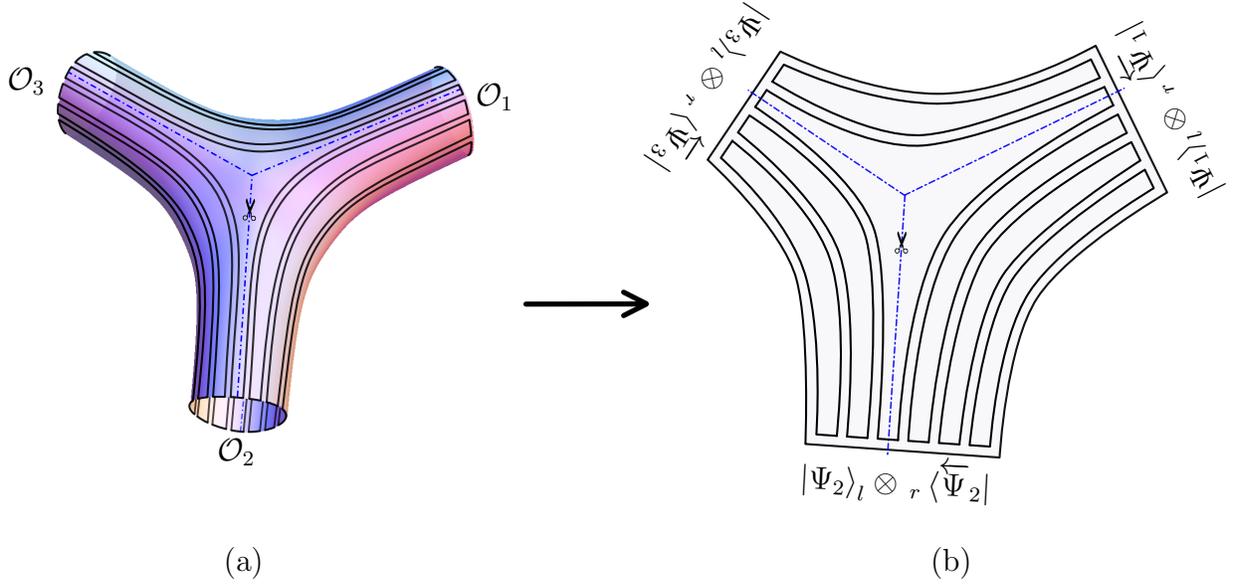
\else
    \com{PDF-picture replacement}
\fi

\caption{\small (a) The planar tree-level contraction of three single trace operators in the double line notation. The diagram has a pair of pants topology (sphere with three punctures). (b) In the spin chain picture, each of the three single trace operators corresponds to a state on a closed chain. The closed chains are cut into right and left open chains where the external states are represented. The three states are sewed together into the three-point function by overlapping the wave functions on each right chain with the wave function on the left subchain of the next operator.}\label{3D3P}
\end{figure}

We are therefore lead to the following picture (see figure \ref{3D3P}a). The three single trace operators of lengths $L_1,L_2$ and $L_3$ are contracted by free propagators. Since each propagator connects two fields, $L_1+L_2+L_3$ must be an even number. The number of free contractions between $\O_i$ and $\O_j$ is  $(L_i+L_j-L_k)/2$. These propagators automatically reproduce the factor $1/|x_{ij}|^{\Delta_i+\Delta_j-\Delta_k}$ in (\ref{3pt}), with $\Delta_i=$ the free dimension $\Delta_i^{(0)}$. In the planar limit, these propagators are all color neighbors. The tree-level structure constant is then given by the sum over all such contractions, normalized by the two-point functions. This paper is devoted to the study of this interesting combinatorial problem.

The three operators have definite one-loop anomalous dimension and are given by a linear combinations of single trace operators. Linear combinations of single trace operators can be represented as states $|\Psi_i\>$ on a closed spin chain. For example, the member of the Konishi multiplet composed from two complex scalars $Z$ and $X$ can be represented in a spin half chain by the state
\beq
K\propto \Tr(ZZXX)-\Tr(ZXZX) \equiv \left|  \uparrow\uparrow \downarrow  \downarrow\right\>-\left|  \uparrow\downarrow \uparrow \downarrow \right\> \, .\la{Konishi}
\eeq
In the spin chain language, the structure constant is constructed by going through the following steps (see also figure \ref{3D3P}b)
\begin{enumerate}
\item Starting with three closed chains in the states $|\Psi_i\>,|\Psi_j\>,|\Psi_k\>$ and choosing a cyclic ordering $(i,j,k)$,
\item Breaking the $i$th closed chain into {\it left} and {\it right} open subchains of lengths $(L_i+L_j-L_k)/2$, $(L_i+L_k-L_j)/2$ and doing the same for the other two closed chains,
\item Expressing the closed chain state as an entangled state in the tensor product of the two subchains Hilbert spaces with the lengths indicated in the previous point. I.e., $|\Psi_i\>=\sum_a |\Psi_{i_a}\>_l \otimes |\Psi_{i_a}\>_r$,
\item  We want to Wick contract the operator corresponding to the state $|\Psi_i\>_r$ with the operator corresponding to the state $|\Psi_{i+1}\>_l$. The Wick contraction is obtained from the spin chain contraction of a ket and bra states $\,_r\< \overleftarrow{\Psi}_i|\Psi_{i+1}\>_l$ after a \textit{flipping}  operation $|\Psi_i\>_l\otimes |\Psi_i\>_r\ \to |\Psi_i\>_l\otimes\,_r\<\overleftarrow{\Psi}_i|$, see figure \ref{3D3P}b.  This flipping operation maps the ket states in the right subchain into bra states with 1) reversed spin chain sites, 2) same wave function (not conjugated), 3) same charges \footnote{This point might be a bit confusing at first. It is explained in greater detail in section \ref{sectailoring} and was previously considered in \cite{Okuyama:2004bd}. To help the curious reader let us give an example of this map of ket into bra: $\left|ZZXZX\right\rangle \to \left\langle \bar X\bar Z\bar X\bar Z\bar Z \right|=\(|\bar X\bar Z\bar X\bar Z\bar Z\>\)^\dagger \neq \(| ZZXZX\>\)^\dagger $.  Note also that in our notations $\<Z|Z\>=\<\bar Z|\bar Z\>=1$, $\<\bar Z|Z\>=0$ and the spins  in the kets and bras are ordered from left to right, i.e. $\< Z X|ZX\>=1$ while $\< Z X|XZ\>=0$.},
\item
Normalizing the three external states.
\end{enumerate}

The resulting structure constant can then be obtained from brute force contractions of the states and is given by
\beq\la{BroughtForce}
c_{ijk}^{(0)}={L_iL_jL_k\sum\limits_{a,b,c} \ _r\< \overleftarrow{\Psi}_{k_c}|\Psi_{i_a}\>_l\ _r\< \overleftarrow{\Psi}_{i_a}| \Psi_{j_b}\>_l\ _r\< \overleftarrow{\Psi}_{j_b}|\Psi_{k_c}\>_l\over \sqrt{L_i\<\Psi_i|\Psi_i\>}\sqrt{L_j\<\Psi_j|\Psi_j\>}\sqrt{L_k\<\Psi_k|\Psi_k\>}}
\eeq
Here, the factors of $L_i,L_j$ and $L_k$ arise from summing over all the ways of cutting open the closed chains, before gluing them together.

What we have said so far applies to the planar tree-level contraction of any single trace operators, whether they have a definite one-loop anomalous dimension or not.
What we have to do, however, is to consider the operators with definite anomalous dimension.
Generically these operators are some linear combinations of a huge number of single-trace operators.
Even at tree-level the direct calculation of these contractions is a very complicated problem, especially for long operators.
One may hope, however, that the integrability hidden in the dilatation operator may help to simplify this problem.
Indeed, the one-loop anomalous dimension matrix is represented by an integrable spin chain Hamiltonian \cite{MZ}. We can therefore use integrability
techniques to compute the normalization of the external eigenstates, called {\it Bethe eigenstates}. Moreover, at one loop the spin chain Hamiltonian only acts {\it locally} on the chain.
As a result, when we decompose the external state as an entangled state in the two subchains, each of the subchain states still has the same local form as an eigenstate.
A state of that local form is called a {\it Bethe state}. Therefore, even though the subchain states are not eigenstates, we can still use integrability techniques to compute their overlaps. We will show that the integrability techniques applied to this problem
are much more efficient than a brute force calculation
and in particular allow us to make computations for asymptotically long operators, which is otherwise impossible.

 To summarize, the tools we need in order to compute the tree-level structure constants are
 \begin{itemize}
 \item The spin chain Bethe eigenstates.
\item The decomposition of an external eigenstate into an entangled state on the direct product of the two subchains. We shall denote this decomposition procedure by \textit{cutting}.
\item Once the state is cut into two we need to \textit{flip} one of its halves from a ket into a bra, see figure \ref{3D3P}b.
\item The overlaps and norms of Bethe states. We denote the overlapping computations by \textit{sewing}.
\end{itemize}
The necessary tools for the cutting, flipping and sewing of eigenstates into a three-point function are explained in section 3. For completeness, the Bethe Ansatz in $\N=4$ SYM and the construction of  the corresponding wave functions is reviewed in section 2. The three-point functions are built using the three steps mentioned above in section 4. In this section we also discuss some interesting limits such as the BMN limit or the classical limit. Section \ref{concl} contains the conclusions and open problems. Appendices A,B,C,D contain supplementary details while appendix E contains examples of structure constants at one loop order and is intended to be self-contained.

\section{Bethe Ansatz. Review and Notation} \la{section2}
In this section we will review some basic facts about two-point functions in $\mathcal{N}=4$ SYM, the map between operators and quantum spin chain states of an integrable spin chain and the coordinate and algebraic Bethe ansatz which can be used to efficiently study the spectrum problem at one loop. One of the goals of this section is to set the notation for the following sections.  The expert reader is certainly familiar with the content of this section and can safely skip it.

In this paper we will mostly consider operators made out of two complex scalars. One such example is the Konishi operator presented in (\ref{Konishi}) made out of the scalars $Z$ and $X$. As mentioned above, we can represent such operators as states in an $SU(2)$ spin chain. One of the scalars is thought of as being a spin up -- the $Z$ field in the example (\ref{Konishi}) -- while the other scalar is thought of as being a spin down --  the $X$ field in the example (\ref{Konishi}). More generally we represent the operator made of $L$ scalar fields $Z$ by a ferromagnetic vacuum state of $L$ spins up \cite{BMN}. Operators with $N$ scalar fields $X$ and $L-N$ fields $Z$ are represented by flipping $N$ of those spins. These spin flip excitations are called magnons. In the example (\ref{Konishi}) we have $L=4$ and $N=2$. A generic state will be of the form
\beq
|\Psi\> = \sum_{1 \le n_1<n_2<\dots < n_N\le L} \psi(n_1,\dots,n_N) |n_1,\dots , n_N\>  \la{psiCBA}
\eeq
where the ket $|n_1,\dots,n_N\>$ stands for the state with spins down at positions $n_1, n_2, \dots n_N$\footnote{E.g., $ \Tr\(Z^{n_1-1} X Z^{n_2-n_1-1} X Z^{n_3-n_2-1} X Z^{L-n_3}\)\mapsto|n_1,n_2,n_3\>$} while $\psi(n_1,\dots, n_N)$ is the wave function, which we will fix below. In what follows, whenever we will refer to a wave function, we will mean the wave function in this local base.

The operators with definite anomalous dimensions are  the eigenvectors of the one-loop mixing matrix $\hat H$ (by definition). The anomalous dimensions are the corresponding eigenvalues (again by definition).
This mixing matrix can be computed in perturbation theory. In the spin chain language, the mixing matrix is represented by a \textit{local} spin chain Hamiltonian of the form \cite{MZ}
\beq
\hat H={\lambda \over 8\pi^2} \sum_{n=1}^L \(\mathbb{I}_{n,n+1}-\mathbb{P}_{n,n+1}\). \la{Hsu2}
\eeq
The permutation operator $\mathbb{P}_{n,n+1}$ acts on the spins at positions $n$ and $n+1$ swapping them. The identity $\mathbb{I}_{n,n+1}$ does nothing and $L+1\equiv 1$. The energy spectrum of this Hamiltonian gives us  the anomalous dimensions of the operators. We encourage the reader who is not familiar with this language to check that the state (\ref{Konishi}) is indeed an eigenvector of this Hamiltonian with anomalous dimension $\gamma_K=3\lambda/4\pi^2$.

We will now review how to construct the wave functions $\psi(n_1,\dots,n_N)$ which diagonalize the spin chain Hamiltonian (\ref{Hsu2}).  For pedagogical references on the coordinate Bethe ansatz, see \cite{Korepinbook, Gaudinbook, CBAintro}.

\subsection{Coordinate Bethe ansatz}

For a single magnon we diagonalize the translation invariant Hamiltonian by going to Fourier, $\psi(n_1)=e^{i p_1 n_1}$. The energy of this state, also called the dispersion relation, is given by $\gamma=\epsilon(p_1)$ with
\beq
\epsilon(p)={\lambda \over 2\pi^2} \sin^2\frac{p}{2} \,.
\eeq
For two particles we write  $\psi(n_1,n_2)=e^{i p_1 n_1+i p_2 n_2}+S(p_2,p_1)e^{i p_2 n_1+i p_1 n_2}$. The relative coefficient between the two plane waves is the amplitude for incoming momenta $\{p_1,p_2\}$ to be exchanged into $\{p_2,p_1\}$. In other words, it is the two body S-matrix. By explicitly acting with $\hat H$ on this state we can read the energy of the state $\gamma=\epsilon(p_1)+\epsilon(p_2)$ and compute the S-matrix,
\beq
S(p,p')=\frac{\frac{1}{2}\cot\frac{p}{2}-\frac{1}{2}\cot\frac{p'}{2}+i}{\frac{1}{2}\cot\frac{p}{2}-\frac{1}{2}\cot\frac{p'}{2}-i} \, . \la{smatrix}
\eeq
Now, in a 1+1 dimensional elastic scattering process between two identical particles, energy and momenta conservation imply that  the individual momentum at most be exchanged.  Hence the two particle ansatz we made \textit{ought} to work.

For three particles the story is radically different. This is when Integrability starts playing a role. Integrability means that in addition to the momentum $Q_1=\hat P$ and energy $Q_2=\hat H$, there exists a tower of local conserved charges $Q_n$ which commute with the momentum and Hamiltonian. We can introduce an arbitrary complex number $u$ and simply encode all conservation laws in\footnote{After some $n$ the charges are of course not independent. For the Hamiltonian (\ref{Hsu2}) such relation can be easily derived using the algebraic Bethe ansatz (ABA) formalism which we will review in subsection \ref{sec2}.}
\beq
\sum_{n=0}^\infty \[\hat P, Q_n\] u^n=\sum_{n=0}^\infty \[\hat H, Q_n\] u^n=0 \,. \la{charges}
\eeq
This relation has important consequences. For example, it allows us to guess the form of $\psi(n_1,n_2,n_3)$ in (\ref{psiCBA}). The reason is that the existence of the higher conserved charges \textit{does} imply that if we scatter three magnons with momenta $\{p_1,p_2,p_3\}$ they will scatter into some other momenta $\{p_1',p_2',p_3'\}$, which must be related to the original ones by a simple reshuffling. In other words, the scattering is effectively pairwise. For example, for three particles we are thus lead to
\beq
\psi(n_1,n_2,n_3)=e^{i p_1 n_1+i p_2 n_2+i p_3 n_3}+ A\, e^{i p_2 n_1+i p_1 n_2+i p_3 n_3} + A' \,e^{i p_2 n_1+i p_3 n_2+i p_1 n_3} + \dots \la{cba3}
\eeq
where $\dots$ stand for the remaining $3$ possible plane waves. The coefficient $A$ multiplies the plane wave which is obtained from the plane wave with unit coefficient by swapping particles with momentum $p_1$ and $p_2$. Thus
\beq
A=S(p_2,p_1) \,,
\eeq
where $S(p,p')$ is the two-body S-matrix (\ref{smatrix}) derived above.
The coefficient $A'$ is the coefficient of the plane wave which is obtained after a sequence of two momentum exchanges,
\beq
A'=S(p_2,p_1)S(p_3,p_1) \,.
\eeq
The coefficients of the three remaining plane waves are obtained in the same way. The generalization to $N>3$ particles involves $N!$ plane waves whose coefficients follow again the same pattern. Our convention for the normalization of the wave function is that the plane wave with no momenta exchanged has unit coefficient, i.e.
\beq
\psi(n_1,\dots,n_N)=e^{i p_1 n_1+ip_2n_2+\dots+ i p_N n_N}+S(p_2,p_1)e^{i p_2 n_1+ip_1n_2+\dots+ i p_N n_N}+\dots \,. \la{waveN}
\eeq
Of course, this normalization depends on the choice of an ordering of the momenta; more on this in section \ref{comments}.
The observation that the eigenstates of the Heisenberg spin chain are given by such ansatz was the key insight of the seminal work of Hans Bethe \cite{Bethe31}. States of the form \eqref{psiCBA}, whose wave functions are given by (\ref{waveN}) are called \textit{Bethe states}.

The energy of the multiparticle state is the sum of the energy of the individual magnons.
Since the system is put in a finite circle of length $L$, the spectrum is discrete. The periodicity of the wave function imposes a set of $N$ quantization conditions for the $N$ momenta,
\beq
 e^{i p_j L} \prod_{k\neq j}^N S(p_k,p_j) = 1 \la{BAE}
\eeq
which are called Bethe equations.
The physical meaning of this equation is the following: if we carry a magnon with momentum $p_j$ around the circle, the free propagation phase $p_j L$ plus the phase change due to the scattering with each of the other $N-1$ magnons must give a trivial phase. We shall denote by \textit{Bethe eigenstates} those Bethe states whose momenta are quantized as in (\ref{BAE}).  Furthermore, when dealing with single trace operators, we have to take into account the cyclicity of the trace by imposing the zero momentum condition
\beqq
\prod_{j=1}^N e^{i p_j} =1.
\eeqq

Let us stress again: the fact that the simple wave functions described above diagonalize the Hamiltonian (\ref{Hsu2}) is absolutely remarkable and non-trivial.
A generic spin chain Hamiltonian will not lead to an integrable theory, the scattering will not factorize into two-body scattering events and the \text{set} of momenta of multiparticle states will not be conserved. Hence, in general, the problem will be of exponential complexity and the best we can do is diagonalize small spin chains with a computer. The Bethe ansatz reduces this problem to a polynomial one. For example, the spectrum problem is completely solved by the simple set of algebraic equations (\ref{BAE}).

The algebraic Bethe ansatz \cite{Faddeev}, reviewed in the next subsection, provides the explanation for this ``miracle".
We will review why (\ref{charges}) is indeed true and we will recall that the wave function (\ref{waveN}) can be constructed by acting on the ``ferromagnetic" state  with some beautiful nonlocal ``creation" operators $\mathcal{B}(u)$. For example, the state (\ref{psiCBA}) with wave function (\ref{waveN}) is simply given by
\beq
|\Psi\>  =\mathcal{N}  \mathcal{B}(u_1) \mathcal{B}(u_2) \dots \mathcal{B}(u_N) \left|\uparrow \dots \uparrow \right\rangle
\eeq
where $\mathcal{N}$ is some simple known normalization factor which we will write down later. The rapidities $u_j$ are a convenient parametrization of the momenta $p_j$ given by
\beq
u=\frac{1}{2} \cot \frac{p}{2} \qquad , \qquad e^{ip}=\frac{u+i/2}{u-i/2}\,. \la{rapidity}
\eeq
Note that with this parametrization the Bethe equations (\ref{BAE}) become simple polynomial equations
\beq
e^{i \phi_j}=1 \qquad \text{where} \qquad e^{i \phi_j}\equiv \(\frac{u_j+i/2}{u_j-i/2}\)^L  \prod_{k\neq j}^N \frac{u_j-u_k-i}{u_j-u_k+i} \,. \la{BAEu}
\eeq

\subsection{Algebraic Bethe ansatz} \la{sec2}
The algebraic Bethe ansatz is a formalism developed by the Leningrad school for constructively producing and solving\footnote{Originally ``solving" meant ``computing the spectrum".} integrable theories (see \cite{Korepinbook,Faddeev, Slavnovreview} for nice introductions to the subject and for further references). The great advantage over the coordinate Bethe ansatz is the constructive nature of the method and the mathematical elegance. The main drawback is that the physical picture is somehow obscured. For example, the magnon physical picture of (\ref{waveN}) is somehow hidden in this formalism.

Let us start by recalling that the Hamiltonian (\ref{Hsu2}) also follows from the following object:
\beq
\hat H-\frac{ \lambda L }{8\pi^2}=-{\lambda \over 8\pi^2} \sum_{n=1}^L  \mathbb{P}_{n,n+1} =\frac{ \lambda}{8i\pi^2}  \frac{d}{du} \Big( \log \Tr_{0}\[ \(u\, \mathbb{I}+i\, \mathbb{P} \)_{01} \dots  \( u\, \mathbb{I}+i\, \mathbb{P} \)_{0L}\] \Big)_{u=0} \la{HABA} \,.
\eeq
Before deriving this fact or explaining its relevance, let us digest the notation introduced. First, $u$ is the so-called \textit{spectral parameter}. It is an arbitrary complex number which we set to zero after taking the derivative\footnote{The variable $u$ will be soon identified with $u$ appearing in (\ref{charges}) and (\ref{rapidity}) in the previous section. }. The \textit{R-matrix}
\beq
R_{0j}(u)=\( u\, \mathbb{I}+i\, \mathbb{P}  \)_{0j} \la{Rmatrix}
\eeq
is an operator acting on a tensor product of two vector spaces: the physical spin chain site vector space at site $j$ and an extra auxiliary space labeled by the index $0$ which we introduced. Both these spaces are isomorphic to $\mathbb{C}^2$, the space where the spin $1/2$ lives. As before, $\mathbb{P}$ is the permutation operator that permute the spins at physical position $j$ and auxiliary position $0$. We can of course write the R-matrix as a simple $4\times 4$ matrix acting on the vector space $\left|\uparrow\uparrow\right\rangle,\left|\uparrow  \downarrow \right\rangle,\left| \downarrow  \uparrow\right\rangle,\left| \downarrow   \downarrow \right\rangle$.

The operator inside the square brackets in (\ref{HABA}) is the \textit{monodromy matrix}
\beq
L_0(u)=R_{01}(u)\dots R_{0L}(u) \,. \la{LR}
\eeq
It acts on the tensor product of $L+1$ spaces: the $L$ physical spaces corresponding to the $L$ lattice sites plus the auxiliary space $0$.  Since it is crucial to understand well the notation we are introducing let us be maximally pedestrian for a second. Using indices $i_i,\dots i_L$, $j_1,\dots,j_L$ for the physical spaces and $a,b$ for the auxiliary space, the monodromy matrix is an object with indices
\beqq
L_0(u)_{i_1,\dots i_L;a}^{j_1\dots j_L;b} \equiv
\( \begin{array}{cc} L_0(u)_{i_1,\dots i_L;1}^{j_1\dots j_L;1} & L_0(u)_{i_1,\dots i_L;1}^{j_1\dots j_L;2} \\
L_0(u)_{i_1,\dots i_L;2}^{j_1\dots j_L;1} & L_0(u)_{i_1,\dots i_L;2}^{j_1\dots j_L;2}
\end{array}\)
\equiv \( \begin{array}{cc} \mathcal A(u+i/2)_{i_1,\dots i_L}^{j_1\dots j_L} & \mathcal B(u+i/2)_{i_1,\dots i_L}^{j_1\dots j_L} \\
\mathcal C(u+i/2)_{i_1,\dots i_L}^{j_1\dots j_L} & \mathcal D(u+i/2)_{i_1,\dots i_L}^{j_1\dots j_L}
\end{array}\)^{b}_{a} \,.
\eeqq
The shifts by $i/2$ are introduced for future convenience.  In other words, we can make the $\mathbb{C}^2$ auxiliary space manifest and write
\beq
L_0(u)=\(\begin{array}{cc}
\mathcal A(u+i/2) & \mathcal B(u+i/2) \\
\mathcal C(u+i/2) & \mathcal D(u+i/2)
\end{array}\)
\eeq
where $\mathcal A, \mathcal B, \mathcal C, \mathcal D$ are operators which only have physical indices, i.e.\ they act on the physical spin chain Hilbert space $\mathcal{H}=(\mathbb{C}^2)^{\otimes L}$.

\begin{figure}[t]
\centering
\def\svgwidth{16cm}

\ifpdf
    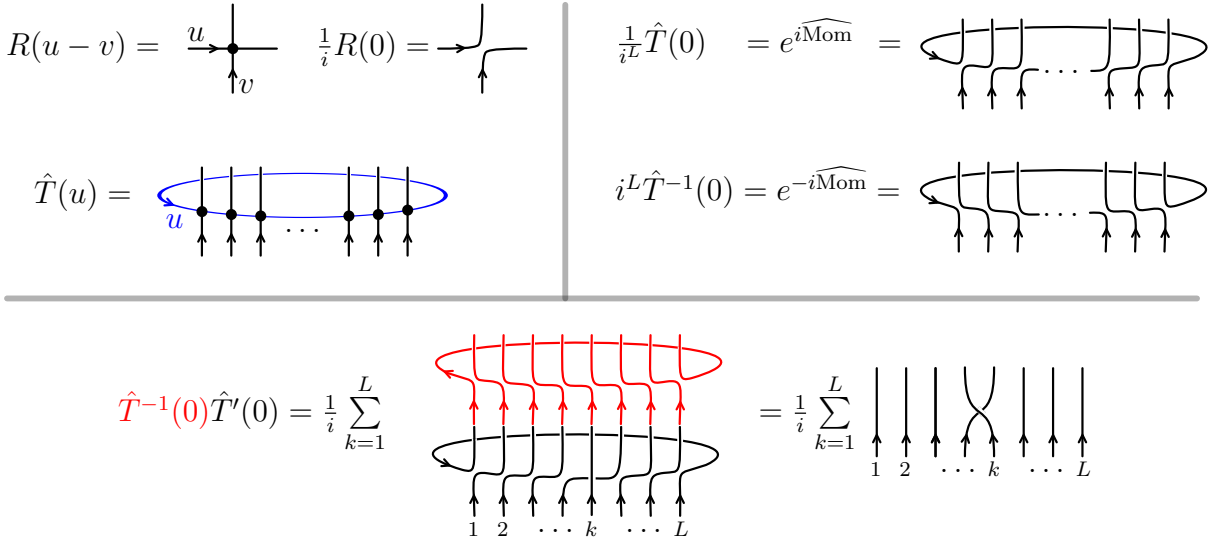
\else
    \com{PDF-picture replacement}
\fi

\caption{\small Since $i^{-1}R(0)=\mathbb{P}$, $i^{-L}\hat T(0)$ is the unit shift operator to the right, see the upper right corner of the figure. By definition of inverse $i^{L}\hat T^{-1}(0)$  is the unit shift operator to the left. When computing $\hat T'(0)$ the derivative will act on one of the $R$'s at position $k$, hence the sum in the last line.  $R'(0)=\mathbb{I} $ which leads to the last line. We see that $\hat T'(0)$ is a sum of terms which are  \textit{almost} a total shift of one unit to the right except for a small ``impurity" at site $k$. Therefore, when multiplying by $\hat T^{-1}(0)$, we \textit{almost} get the identity operator acting on the full Hilbert space. The impurity simply leads to a permutation of acting on sites $k$ and $k-1$. Hence (\ref{HABA2}) leads to  (\ref{Hsu2}). }
\label{derivationH}
\end{figure}

The next object we see in (\ref{HABA}) is the trace with respect to the auxiliary space of the monodromy matrix. This defines the \textit{transfer matrix}
\beq
\hat T(u)\equiv \Tr_0 L_0(u) \,.
\eeq
Since we traced over the auxiliary space the transfer matrix is an operator acting on the physical Hilbert space. More explicitly we can write it in terms of the operators $\mathcal A$ and $\mathcal D$ as
\beq
\hat T(u)= \mathcal A(u+i/2)+ \mathcal D(u+i/2)\,.
\eeq
Finally, in (\ref{HABA}) we have the derivative of the logarithm of an operator (or big matrix). This is understood as usual as the inverse of the matrix times its derivative. With the notation introduced above, (\ref{HABA}) simply reads
\beq
\hat H-\frac{ \lambda  L}{8\pi^2}={\lambda\over 8i\pi^2} \,\hat T^{-1}(0) \,\hat T'(0)  \la{HABA2}
\eeq
where the prime stands for derivative with respect to the spectral parameter $u$. The derivation of (\ref{HABA2}) is illustrated in figure \ref{derivationH}.
\begin{figure}[t]
\centering
\def\svgwidth{14.5cm}

\ifpdf
    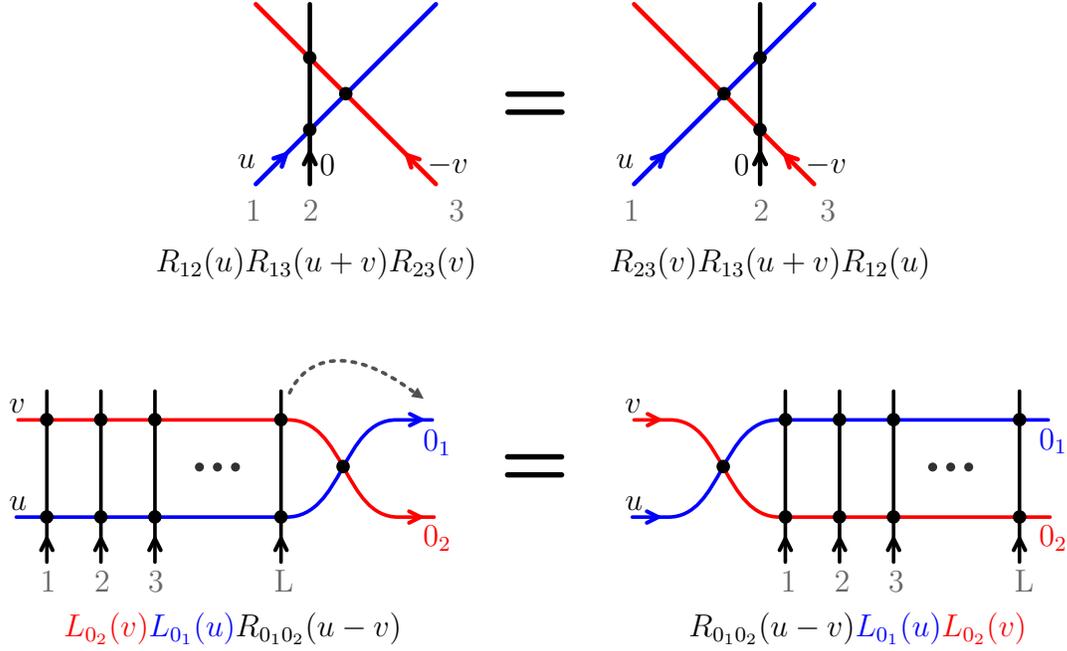
\else
    \com{PDF-picture replacement}
\fi

\caption {\small The R-matrix is a very special operator. It is designed to satisfy the Yang-Baxter equation depicted at the top. This equation is arguably the most important equation in quantum integrability. In particular it implies the $LLR=RLL$ type relation represented at the bottom. To prove this relation we simply move one of the vertical lines from the left group to the right region using Yang-Baxter and repeat this procedure until all the vertical lines are to the right of the R-matrix. From this simple equation all the algebra relations (table \ref{FZalg}) of the monodromy matrix elements $\mathcal A,\mathcal B,\mathcal C,\mathcal D$ follow trivially as explained in the main text. Finally multiplying this equation by $R^{-1}$ and taking the trace over the tensor product of the two auxiliary spaces $0_1$ and $0_2$ we derive (\ref{ttcom}).}
\label{YBfig}
\end{figure}

An important feature of this construction is that, as explained in figure \ref{YBfig}, transfer matrices with different spectral parameters commute,
\beq
\[\hat T(v),\hat T(u)\]=0 \qquad \forall \qquad u,v \la{ttcom} \, .
\eeq
The Hamiltonian is just the log derivative of the transfer matrix and hence $[\hat H,\hat T(u)]=0$. Also $[\hat H,\log \hat T(u)]=0$. The latter relation is quite nice since it implies (\ref{charges}) neatly. Recall that the existence of the higher local charges $Q_n$ is what ensures Integrability of the model.  Indeed, we can Taylor expand $\log \hat T(u)$ around $u=0$,
\beq
\log \hat T(u)=\sum_{n=0}^\infty  Q_n u^n \,,
\eeq
and generate in this way a tower of local\footnote{Of course we would also get conserved charges by expanding $\hat T(u)$ or \textit{any} functional of the transfer matrix around \textit{any} point $u=u^*$. The advantage of expanding $\log \hat T(u)$ around $u=0$ is that the charges generated in this way are local. This follows from the important relation $R(0)=\mathbb{P}$, see the derivation in figure \ref{derivationH} for an illustration of the importance of this property.} conserved charges. The first one, $Q_1$, is the momentum and  the second, $Q_2$, is the energy, see figure \ref{derivationH}. The higher terms are the higher charges we were looking for. Hence, indirectly, we now understand why the multi-particle ansatz (\ref{waveN}) ought to work.

Relation (\ref{ttcom}) has another interesting consequence. Since the transfer matrices commute with each other for different values of the spectral parameter we can diagonalize $\hat T(u)$ with a spectral parameter \textit{independent} base of $2^N$ states $|\Psi_i\>$ such that
\beq
\hat T(u) |\Psi_i\> =\Big(\mathcal A(u+i/2)+ \mathcal D(u+i/2)\Big)|\Psi_i\> = T(u)_i |\Psi_i\>  \,. \la{TPsi}
\eeq
The states diagonalize all the higher charges at the same time since they are simple derivatives of the transfer matrix. We could have picked any of the higher charges as the Hamiltonian. The algebraic Bethe ansatz (ABA) approach tells us that all these Hamiltonians can be diagonalized at once with the same wave functions! We start seeing the constructive nature of the ABA approach.

We already encountered the state $|\Psi\>$ in the coordinate Bethe ansatz approach: it is simply \eqref{psiCBA} with the wave function given by (\ref{waveN}). For us it is important to construct this state in the ABA language. It turns out that the monodromy matrix elements $\mathcal B(u)$ can be used as creation operators for the magnons.\footnote{This will be made more clear below, see discussion after (\ref{largev}). } More precisely, the claim is that
\beq
|\Psi\>= \mathcal{B}(u_1) \dots \mathcal{B}(u_N) \left|\uparrow \dots \uparrow \right\rangle \, , \la{psiABA}
\eeq
yields a state proportional to \eqref{psiCBA} once we identify $u_i$ and $p_i$ according to (\ref{rapidity}). In other words $|\Psi\>$ is a \textit{Bethe state}.
If $u_i$ are Bethe roots satisfying the Bethe equations (\ref{BAEu}), then $|\Psi\>$ satisfies (\ref{TPsi}) and is a \textit{Bethe eigenstate}.  We will sometimes denote \eqref{psiABA} by $\left| \{u_j\} \right\>^\al$.

To understand why (\ref{TPsi}) indeed holds we need to 1) understand how to take the operators $\mathcal A$ and $\mathcal D$ through the creation operators $\mathcal B$ and 2) derive the action of $\mathcal A$ and $\mathcal D$ on the ferromagnetic vacuum.  The latter is simple. From the definition (\ref{LR}) we easily see that
\beqa
&&\mathcal A(u) \left|0\right\rangle=a(u) \left|0\right\rangle \,\,\, , \qquad \mathcal D(u) \left|0\right\rangle=d(u) \left|0\right\rangle \, , \la{ADzero}\\
&&\mathcal C(u) \left|0\right\rangle=0\,\,\,\,\, \qquad \,\,\,\, , \,\qquad \mathcal B(u) \left|0\right\rangle=\text{a non-trivial single spin excitation} \, ,
\la{actionABCD}
\eeqa
where $\left|0\right\rangle=\left|\uparrow \dots \uparrow \right\rangle$ and\footnote{Different overall normalizations of the $R$ matrix (\ref{Rmatrix}) lead to different functions $a(u)$ and $d(u)$. The ration $e(u)=a(u)/d(u)$ is however independent of that choice of normalization and any physical quantity will depend in $a$ and $d$ only through that ratio. With our choice of normalization for the $R$-matrix we have $\mathcal{B}^{\dagger}(u)=-\mathcal{C}(u)$. }
\beq
a(u) \equiv \({u+\frac{i}{2}}{}\)^L  \qquad , \qquad d(u) \equiv \({u-\frac{i}{2}}{}\)^L \,.
\la{ad}
\eeq
Next we need to understand the algebra of the elements of the monodromy matrix $\mathcal A,\mathcal B,\mathcal C$ and $\mathcal D$.
As explained in figure \ref{YBfig} the monodromy matrix $L(u)$ satisfies the relation
\beq
L_{0_2}(v)L_{0_1}(u)R_{0_10_2}(u-v) = R_{0_10_2}(u-v)L_{0_1}(u)L_{0_2}(v) \,,
 \la{RLL}
\eeq
where $0_1$ and $0_2$ are two auxiliary spaces isomorphic to $\mathbb{C}^2$. This relation encodes the algebra of the monodromy matrix elements. More explicitly
we can choose a basis $\left|\uparrow\uparrow\right\rangle,\left|\uparrow  \downarrow \right\rangle,\left| \downarrow  \uparrow\right\rangle,\left| \downarrow   \downarrow \right\rangle$ spanning the tensor product of these two spaces. In this basis the left-hand side of (\ref{RLL}) simply reads
\beq
\begin{pmatrix}
\mathcal A(v) & \mathcal B(v) & 0 & 0 \\
\mathcal C(v) & \mathcal D(v) & 0 & 0 \\
0 & 0 & \mathcal A(v) & \mathcal B(v) \\
0 & 0 & \mathcal C(v) & \mathcal D(v)
\end{pmatrix} \! \! \nn
\begin{pmatrix}
\mathcal A(u) & 0 & \mathcal B(u) & 0 \\
0 & \mathcal A(u) & 0 & \mathcal B(u) \\
\mathcal C(u) & 0 & \mathcal D(u) & 0 \\
0 & \mathcal C(u) & 0 & \mathcal D(u)
\end{pmatrix} \! \!
\begin{pmatrix}
u\!-\!v\!+\!i & 0 & 0 & 0 \\
0 &u\!-\!v &{i} & 0 \\
0 &{i}  &u\!-\!v& 0 \\
0 & 0 & 0 &u\!-\!v\!+\!i
\end{pmatrix}
\eeq
while the right-hand side is obtained by multiplying the same matrices in the opposite order.  In this way we get $16$ algebra relations between the several elements which we summarize in table \ref{FZalg}.

\begin{table}[t]
\begin{align}
\mathcal A(v) \mathcal B(u) &= f(u-v) \mathcal B(u) \mathcal A(v) + g(v-u) \mathcal B(v) \mathcal A(u) \,  \la{AB} \\
\mathcal B(v) \mathcal A(u) &= f(u-v) \mathcal A(u) \mathcal B(v) + g(v-u) \mathcal A(v) \mathcal B(u) \,   \nn \\
\mathcal D(v) \mathcal B(u) &= f(v-u) \mathcal B(u) \mathcal D(v) + g(u-v) \mathcal B(v) \mathcal D(u) \,  \la{DB} \\
\mathcal B(v) \mathcal D(u) &= f(v-u) \mathcal D(u) \mathcal B(v) + g(u-v) \mathcal D(v) \mathcal B(u) \,  \nn \\
\mathcal C(v) \mathcal A(u) &= f(v-u) \mathcal A(u) \mathcal C(v) + g(u-v) \mathcal A(v) \mathcal C(u) \,  \nn \\
\mathcal A(v) \mathcal C(u) &= f(v-u) \mathcal C(u) \mathcal A(v) + g(u-v) \mathcal C(v) \mathcal A(u) \,  \nn \\
\mathcal C(v) \mathcal D(u) &= f(u-v) \mathcal D(u) \mathcal C(v) + g(v-u) \mathcal D(v) \mathcal C(u) \,  \nn \\
\mathcal D(v) \mathcal C(u) &= f(u-v) \mathcal C(u) \mathcal D(v) + g(v-u) \mathcal C(v) \mathcal D(u) \,  \nn \\
\[\mathcal C(v),\mathcal B(u)\] &= g(u-v) \[\mathcal A(v) \mathcal D(u) - \mathcal A(u) \mathcal D(v)\]= g(u-v) \[\mathcal D(u) \mathcal A(v) - \mathcal D(v) \mathcal A(u)\] \la{CB} \\
\[\mathcal D(v), \mathcal A(u)\] &= g(u-v) \[\mathcal B(v) \mathcal C(u) - \mathcal B(u) \mathcal C(v)\]= g(u-v) \[\mathcal C(u) \mathcal B(v) - \mathcal C(v) \mathcal B(u)\]  \nn\\
\[\mathcal B(u), \mathcal B(v)\] &=
\[\mathcal C(u), \mathcal C(v)\] =
\[\mathcal A(u), \mathcal A(v)\] =
\[\mathcal D(u), \mathcal D(v)\] =0 \la{BB}
\end{align}
where
\beq
g(u)\equiv \frac{i}{u} \qquad , \qquad f(u) \equiv 1+ \frac{i}{u} \,.
\la{fg}
\eeq
\caption{Algebra of the monodromy matrix elements which follows from the $RLL=LLR$ relations described in the text and depicted in figure \ref{ttcom}.}
\la{FZalg}
\end{table}
To justify the identification of $\mathcal B$ and $\mathcal C$ with creation and annihilation operators, we compute the commutation relations of these nonlocal operators with the total spin generators $S^i \equiv \sum_{n=1}^L \sigma^i$. For that aim we consider the limit $v\to \infty$ with $u$ held fixed. In this limit we have, from the definition (\ref{LR}), that $ \mathcal A(v)/a(v) \simeq 1+ \frac{i}{2v} \(1+S^z\)$, $ \mathcal C(v)/a(v)\simeq \frac{i}{v} S^+$, $\mathcal D(v)/a(v) \simeq 1+ \frac{i}{2v} \(1-S^z\)$ and
\beq
\frac{\mathcal B(v)}{a(v)} \simeq \frac{i}{v} \, S^-\ . \la{largev}
\eeq
Now, from the sixteen algebra relations in table \ref{FZalg} we can read for example the commutation relations $ \[\mathcal B(u),S^z\]=2 \, \mathcal B(u) $ and $ \[\mathcal C(u),S^z\]= -2 \, \mathcal C(u)$ which means that $\mathcal B(u)$ is a single spin creation operators while $\mathcal C(u)$ is a single spin annihilation operator. We also have $\[\mathcal A(u),S^z\]=\[\mathcal D(u),S^z\]=0$. Furthermore,
\beq
\[ S^+,\mathcal B(u) \]=\mathcal A(u)-\mathcal D(u) \,\, ,  \la{BSp}
\eeq
Equations (\ref{largev}) and (\ref{BSp}) have further important consequences which we will discuss at the end of the next subsection.

Using relations (\ref{AB}) and (\ref{DB}) we learn how to carry $\mathcal A$ and $\mathcal D$ through the $\mathcal B$ operators.
Suppose we would drop the second term in the right-hand side of (\ref{AB}) and (\ref{DB}). Then it is clear that (\ref{psiABA}) would be an eigenstate of $\hat T(u-i/2)$ with eigenvalue
\beq
T(u-i/2)=\({u+\frac{i}{2}}{}\)^L \prod_{j=1}^N \frac{u-u_j-i}{u-u_j} + \({{u-{i\over2}}}{} \)^L  \prod_{j=1}^N \frac{u-u_j+i}{u-u_j}   \, .
\la{Teigenval}
\eeq
Because of the second term in the right hand side of (\ref{AB}) and (\ref{DB}) the state $|\Psi\>$ is typically \textit{not} an eigenstate of the transfer matrix. We can wonder what  conditions do we need to impose so that the contribution of these extra terms vanishes. These conditions are nothing but the Bethe equations (\ref{BAEu})!\footnote{\la{foot}
Indeed, those extra terms would lead to terms of the form
$
\sum_{k=1}^N (\alpha_k+\delta_k) \, \mathcal B(u) \prod_{j\neq k}\mathcal{B}(u_j)  \left| \uparrow \dots \uparrow \right\>
$ where $\alpha_k$ is the contribution coming from $\mathcal A(u)$ and $\delta_k$ is the contribution coming from $\mathcal D(u)$ in (\ref{TPsi}). Suppose we want to find $\alpha_k$. We start with
$\mathcal A(u) \mathcal{B}(u_k) \prod_{j\neq k} \mathcal{B}(u_j) \left| \uparrow \dots \uparrow \right\>$ and need to carry $\mathcal A(u)$ to the right till it hits the vacuum. Note that we ordered the creation operators so that $\mathcal B(u_k)$ comes first. We can always do it since $\[\mathcal B(u),\mathcal B(v)\]=0$. Now it is clear that to end up with terms contributing to $\alpha_k$ we must first use the second term in (\ref{AB}) to commute $\mathcal A(u)$  through $\mathcal{B}(u_k)$ so that $\mathcal{B}(u_k)$ becomes $\mathcal B(u)$ ($\mathcal A(u)$ becomes $\mathcal A(u_k)$). \textit{But then} we must  always use the first term in (\ref{AB}) to commute $\mathcal A(u_k)$ through all other $\mathcal B$'s since we are already happy with the arguments of the $\mathcal B$ operators. Hence we conclude that $\a_k = a(u_k) g(u-u_k) \prod_{j \neq k}^N f(u_j-u_k)$. Similarly $\delta_k = d(u_k) g(u_k-u) \prod_{j \neq k}^N f(u_k-u_j) $.  The condition $\alpha_k+\delta_k=0$ yields the Bethe equations (\ref{BAEu}) once we recall (\ref{fg}) and (\ref{ad}).}

\subsection{A few comments and summary of notation} \la{comments}
As explained above the Bethe states \eqref{psiCBA} and (\ref{psiABA}) are proportional to each other. Let us write a precise relation between the two.
First recall that to define the coordinate Bethe state normalized according to (\ref{waveN}) we pick a particular order for the momenta. Given an ordering $p_1,\dots,p_N$ for the momenta, the relation between the Bethe states normalized according to the coordinate Bethe ansatz (CBA) (\ref{waveN}) and the algebraic Bethe ansatz (ABA) (\ref{psiABA}) is\footnote{In the local basis of states (\ref{psiCBA}), the wave functions are rational function of the rapidities $\{u_i\}$. The conversion factor in (\ref{conversion}) can be simply derived by demanding that the two wave functions have the same zeros, poles and large $u_k$ behavior.}
\beq
|\Psi\>^{\al} =  d^{\{u\}} \, g^{\{u+\frac{i}{2}\}}f_<^{\{u\}\{u\}}\ |\Psi\>^{\co} \, ,
\la{conversion}
\eeq
where we have introduced the shorthand notation
\beq
F^{\{u\}} \equiv \prod_{u_j\in \{u\}} F(u_j)\;\;,\;\;F_<^{\{u\}\{u\}} \equiv\!\!\!\!\!\!\! \prod_{\scriptsize\bea{c}u_i,u_j\in \{u\}\\i<j\eea}\!\!\!\!\!\!\! F(u_i-u_j)
\la{prodnot}
\eeq
which we will use throughout the rest of the paper.  Also, from now on we should explicitly attach an upper index ``al'' or ``co'' where necessary to specify if we are referring to a given object in the algebraic or coordinate normalizations.

Of course the normalization of the algebraic Bethe ansatz state (\ref{psiABA}) does not depend on the order of the momenta. After all, we see from \eqref{BB} that the $B(u_i)$ operators commute with each other. Hence we can also use (\ref{conversion}) to go between two coordinate Bethe state with different ordering of the same magnons.

Let us re-write the Konishi state (\ref{Konishi}) in the coordinate and algebraic normalizations. The Konishi state has $L=4$, $N=2$ and momenta $p_1=-p_2=2\pi/3$ or $u_1=-u_2={\sqrt{3}}/{6}$. This leads to
\begin{align}
K &= 4 \,e^{-i \pi/3}  \Bigl[ \Tr(ZZXX) - \Tr(ZXZX) \Bigr] \, , & &\text{in the CBA normalization,} \nn\\
K &= \frac{8}{27}  \Bigl[ \Tr(ZZXX) - \Tr(ZXZX)  \Bigr] \, , & &\text{in the ABA normalization.} \nn
\end{align}

Finally let us end this section with a discussion on the completeness of the basis of states discussed so far. Using (\ref{BSp}) and the same kind of reasonings of footnote \ref{foot} we can easily prove that
\beq
S^+ \mathcal{B}(u_1) \dots \mathcal{B}(u_N) \left|\uparrow \dots \uparrow \right\rangle =0
\eeq
if the roots $u_j$ obey Bethe equations. In other words, $S^+$ kills Bethe \textit{eigenstates}. This means that Bethe eigenstates are \textit{highest weight} states. Indeed, if we count the solutions to the Bethe equations (\ref{BAEu}) we find that there are precisely \cite{Faddeev}
\beqq
Z_{L,N}={L \choose N} - {L \choose N-1}
\eeqq
solutions with $N$ spin flips. This is exactly the number of highest weights for a chain of length $L$ and $S^z=L/2-N$. All other states are found by acting with $S^-$ on these states. Indeed
 \beq
 \sum_{N=0}^{L/2} \[2\(\frac{L}{2}-N\)+1\] Z_{L,N}=2^L \,.
 \eeq
Hence a complete basis of states is given by
\beq
\(S^-\)^n|\{u_j\}\>^\al=\(S^-\)^n \mathcal{B}(u_1) \dots \mathcal{B}(u_N) \left|\uparrow \dots \uparrow \right\rangle
\eeq
where $u_j$ obey Bethe equations and $n=0,\dots,L-2N$. All these states have the same energy (or any other charge) as the highest weight state $\mathcal{B}(u_1) \dots \mathcal{B}(u_N) \left|\uparrow \dots \uparrow \right\rangle $ since the Hamiltonian (or the transfer matrix) commutes with $S^i$ for $i=z,+,-$.

Note also that these states can be written using only the creation operators $\mathcal{B}(u)$ since for large rapidities this operator becomes the lowering operator, see (\ref{largev}). More precisely, this state is simply given by
\beq
\(S^-\)^n|\{u_j\}\>^\al=\lim_{\Lambda_{k}\to \infty} \(\prod_{k=1}^n \frac{\Lambda_k^{1-L}}{i}\) \mathcal{B}( \Lambda_1) \dots \mathcal{B}( \Lambda_n) \mathcal{B}(u_1) \dots \mathcal{B}(u_N) \left|\uparrow \dots \uparrow \right\rangle \, .
\eeq
In short, we see that if we consider the Bethe equations for $u_j$ and then add a few Bethe roots at infinity we describe the full Hilbert space.

The factor $ \(\prod_{k=1}^n \Lambda_k^{1-L}/i\)$ simply ensures a good limit when the roots go to infinity. The coordinate Bethe ansatz states have a better limit when we send some of the roots $u_*$ to infinity
\beq\la{descendant}
\lim_{u_k\to\infty}|\{u_j\}\>^\co=S^-|\{u_j\}_{j\ne k}\>^\co \, .
\eeq
Particularly important states are the so-called vacuum descendants
\beq
|\{\infty^N\}\>^\co=\(S^-\)^N \left| \uparrow \dots \uparrow \right\rangle
\eeq
which correspond to a state with $N$ roots at infinity only. They correspond to a ferromagnetic vacuum rotated away from the $z$ axis. In the $\mathcal{N}=4$ language it corresponds to an operator with $L-N$ scalar fields $Z$ and $N$ scalar fields $X$ without any anomalous dimension. These are BPS states whose anomalous dimension is zero to all orders in perturbation theory as a consequence of supersymmetry. In the string theory dual language these states correspond to BMN point like strings which rotate around one of the equators of $S^5$ at the speed of light. Different values of $N$ correspond to different equators of $S^5$ which are of course related by a global rotation.

\section{Tailoring tools} \la{sectailoring}
In this section we will present the necessary tools for computing the three-point function in the fashion explained in the introduction. That is, we will express a closed chain Bethe eigenstate as an entangled state in the tensor product of the two open subchains Hilbert spaces (subsection \ref{cutting:sec}). Then we will explain the flipping procedure, mapping a state in one of the two subchains from ket to bra (subsection \ref{flipping:sec}). Finally, we will compute the overlap of Bethe states (subsection \ref{sewing:sec}). We denote these three procedures by \textit{cutting}, \textit{flipping} and \textit{sewing}.   In section \ref{four} these tools will be used to put the pieces together into the structure constant as illustrated in figure \ref{3D3P}.  Two references that give a nice, pedagogical introduction to the techniques used in the cutting and sewing sections below (from the algebraic basis point of view) are \cite{Korepinbook, Slavnovreview}.

\subsection{Cutting...} \la{cutting:sec}
Consider a spin chain of length $L$ and a generic state (\ref{psiCBA}) of that spin chain. We denote the first $l$ spins starting from the left of that chain by \textit{left} subchain and the last $r=L-l$ spins by \textit{right} subchain.
We can represent that state in the big chain (\ref{psiCBA}) as an entangled state in the tensor product of the left and right subchains as
\beq\la{Break}
 |\Psi\>=\sum_{k=0}^{\min\{N,l\}} \sum_ {1\le n_1<\dots <n_k\le l} \sum_{l<n_{k+1}<\dots < n_N}\!\!\!\!\!\!\! \psi(n_1,\dots,n_N)\,  |n_1,n_2,\dots,n_k\>\otimes |n_{k+1}-l,\dots,n_N-l\> \, .
\eeq
The sum over $k$ is the sum over how many magnons are in the left chain.
The sum on the right hand side of (\ref{Break}) has ${L\choose N}$ terms. For Bethe states  there is a huge simplification. Namely, when we represent a Bethe state as an entangled state in the two subchains, each of the subchain states still has the same Bethe state form.

The reason is that -- locally -- these are eigenstates of a local Hamiltonian and therefore take the specific local form presented  in the previous section. As a result, when we represent a Bethe state as an entangled state in the two subchains, each of the subchain states still has the same local Bethe state form.

In other words, a magnon that is locally propagating along the chain does not know that at some far away point the chain was broken.
To write the corresponding piece of the wave function, all we need to know is whether that magnon propagates on the left or right subchains. A Bethe state therefore breaks into two as
\beq\la{BreakBethe}
|\{u_i\}\>=\sum_{\alpha\cup\bar\alpha}H(\alpha,\bar\alpha)|\alpha\>_l\otimes|\bar\alpha\>_r \, ,
\eeq
where the sum is over all  $2^N$ possible way of splitting the rapidities into two groups $\alpha$ and $\bar\alpha$ such that $\alpha\cup\bar\alpha=\{u_i\}$.  For example, if $N=2$, the possible partitions $\( \a,\bar \a \)$ would be $\( \{\},\{u_1,u_2\} \) ,  \( \{u_1\},\{u_2\} \),  \( \{u_2\},\{u_1\} \),  \( \{u_1,u_2\}, \{\} \)$.
As opposed to the ${L\choose N}$ terms in (\ref{Break}), we only have $2^N$ terms in (\ref{BreakBethe}). This is a very convenient simplification whenever $L\gg N$.

The function $H(\alpha,\bar\alpha)$ takes different forms in the coordinate and algebraic bases. Below we will compute that function for each of these normalizations.
Before moving on, let us introduce some further notation in the same spirit of (\ref{prodnot}). We shall use
\beq
F^\alpha \equiv \prod_{u_j \in \alpha} F(u_j)\ ,\qquad F^{\alpha\bar\alpha} \equiv \!\!\! \prod_{\scriptsize \begin{array}{c} {u_i} \in \alpha\\ v_j \in \bar \alpha\end{array}} \!\!\!F(u_i-v_j)\ ,\qquad F_<^{\alpha\alpha} \equiv \!\!\! \prod_{\scriptsize \begin{array}{c} {u_i, u_j} \in \alpha  \\  i<j \end{array}} \!\!\!F(u_i-u_j) \, . \la{morenotation}
\eeq

\subsubsection*{Cutting a coordinate Bethe state}
The normalization of a Bethe state in the coordinate base $|\{u_i\}\>^\co$ depends on the choice of ordering $u_1,u_2,\dots,u_N$ of the magnons, see (\ref{waveN}). For the states in the two subchains $|\alpha\>^\co_l$ and $|\bar\alpha\>^\co_r$ we choose the ordering induced from the ordering of $|\{u_i\}\>^\co$. With these conventions we will now derive $H(\alpha,\bar \alpha)$.

We first consider the case where all the magnons are in the left subchain: $\alpha=\{u_i\}$, $\bar\alpha=\emptyset$. In this case, by construction, the wave function of $|\alpha\>^\co_l$ coincides with the wave function of $|\{u_i\}\>^\co$ and therefore
\beq
H^\co(\{u_i\},\emptyset)=1\, .\nn
\eeq
Now suppose we shift some set of magnons from $\alpha$ to $\bar\alpha$. There will be two factors contributing to $H$. One is the phase shift acquired by the $\bar\alpha$ magnons when translated trough the first $l$ sites\footnote{In other words, this factor arises because we shifted the label of the positions of the magnons of the right chain by $l$ in order to start counting them from $1$ as usual, see (\ref{Break}). }
\beq
\prod_{\bar\alpha}\({\bar\alpha_i+{i\over2}\over \bar\alpha_i-{i\over2}}\)^l\equiv {a_l^{\bar\alpha}\over d_l^{\bar\alpha}}\equiv e_l^{\bar\alpha}\, .\nn
\eeq
where $a_{l}$ ($a_r$) and $d_{l}$ ($d_r$) are defined as in (\ref{ad}) but using the lengths of the left (right) subchain instead of $L$.
The second contribution to $H(\alpha,\bar \alpha)$ is the scattering phase between all pairs of magnons $u_j\in\alpha$ and $u_i\in\bar\alpha$ such that $i<j$
\beq
\prod_{\scriptsize\begin{array}{c} i<j \\ \ u_j\in\alpha,\ u_i\in\bar\alpha \end{array}}\!\!\!\!\!\!{f(u_j-u_i)\over f(u_i-u_j)}
\, .
\eeq
Multiplying the two factors and using the shorthand notation introduced in (\ref{morenotation}), we can write $H^\co(\alpha,\bar\alpha)$ as
\beq
H^\co(\alpha,\bar\alpha)=\frac{a_l^{\bar\alpha}}{d_l^{\bar\alpha}}{f^{\alpha\bar\alpha}f_<^{\bar\alpha\bar\alpha}f_<^{\alpha\alpha}\over f_<^{\{u\}\{u\}}}\ .
\eeq
Finally, note that cutting a descendant of a coordinate state is trivially obtained from (\ref{descendant})
\beq\la{BreakBethedesc}
|\{u_i,\infty^n\}\>^\co=\sum_{\alpha\cup\bar\alpha=\{u\}}H^\co(\alpha,\bar\alpha)\sum_{m=0}^n{n\choose m}\,|\alpha\cup\{\infty\}^m\>_l^\co\,\otimes\,|\bar\alpha\cup\{\infty\}^{n-m}\>_r^\co \, .
\eeq

\subsubsection*{Cutting an algebraic Bethe state}

We can now use the translation between the coordinate and the
algebraic normalizations (\ref{conversion}) to conclude that
\beq\la{BreakAlg}
H^\al(\alpha,\bar\alpha)=f^{\alpha\bar\alpha}\, d_r^\alpha a_l^{\bar\alpha}\ .
\eeq
We can also straightforwardly derive (\ref{BreakAlg}) by writing the
monodromy matrix (\ref{LR}) as a product of the monodromy matrices of
the left and right subchains, $L=\( R_{01}\dots
R_{0l}\)\(R_{0l+1}\dots R_{0L}\)=L_l L_r\,.$ More explicitly, the
monodromy matrix elements are split as
\beq
\begin{pmatrix}
\mathcal A(u) & \mathcal B(u) \\
\mathcal C(u) & \mathcal D(u)
\end{pmatrix}=
\begin{pmatrix}
\mathcal A_l(u) & \mathcal B_l(u) \\
\mathcal C_l(u) & \mathcal D_l(u)
\end{pmatrix}
\begin{pmatrix}
\mathcal A_r(u) & \mathcal B_r(u) \\
\mathcal C_r(u) & \mathcal D_r(u)
\end{pmatrix} \, .
\eeq
In particular, the Bethe state (\ref{psiABA}) can be written as
\beq
\mathcal B({\bf u}) |0\>=\prod_{j=1}^N \Big(\mathcal A_l(u_j)\mathcal
B_r(u_j)+\mathcal B_l(u_j)\mathcal D_r({u_j})\Big)|0\> \, . \la{breaking}
\eeq
We could open the parentheses and get a sum of $2^N$ terms
parametrized by which roots end up in the left spin chain creation
operator $\mathcal{B}_l$. That set of roots is denoted by $\alpha$.
The set of roots ending up in the right spin chain creation operator
$\mathcal{B}_r$ is denoted by $\bar\alpha$. Of course $\{u_j\}=\alpha
\cup \bar \alpha$. This is precisely the sum over partitions in
(\ref{BreakBethe}). To get rid of the $\mathcal{A}$ and $\mathcal{D}$
operator in (\ref{breaking}) we use the commutation relations
in table \ref{FZalg} to commute them through the $\mathcal{B}$'s plus
the actions of these operators on the vacuum of the corresponding
chains as given in (\ref{actionABCD}). In this way one arrives at
(\ref{BreakAlg}). This technique is known as the {\it generalized two-component model} \cite{Izergin:1984tq, Slavnov}.

\subsection{... Flipping ...} \la{flipping:sec}
After splitting the states into two (which we did in the previous subsection), we need to represent the Wick contraction of the elementary fields as a spin chain operation (see figure 1). The Wick contraction operation takes two ket states on two subchains and produce a number. It can be achieved by two steps. We first \textit{flip} one of the two kets into bra and then, in the next section, contract the ket and bra states. The result does not depend on which of the two kets we choose to flip. In what follows we will always flip the state in the right subchain before contracting it with the state on the left subchain of the next operator (as illustrated in figure 1b). As explained in the introduction, the \textit{flipping} procedure is not the usual conjugation. The usual conjugation acts on the corresponding operator by conjugation and therefore flips the order of fields and their charges. On the contrary the flipping procedure $\mathcal{F}$ does not change the operator. As a result, the contraction of a ket state with a flipped bra state is the same as the Wick contraction of the corresponding operators.

In the introduction we added an upper arrow to distinguish $\,_l\<\overleftarrow{\Psi}|={\cal F} \circ |\Psi\>_l$ from  $\,_l\<\Psi|=\(|\Psi\>_l\)^\dagger$.
For example, for real $\phi$ we have
\beqa
\dagger &:&\qquad e^{i\phi} | X Z X Z Z\> \to  \< X Z X Z Z | e^{-i\phi} \, ,  \\
 \mathcal{F} &:&\qquad e^{i\phi} | X Z X Z Z\> \to  \< \bar Z \bar Z  \bar X \bar Z \bar X| e^{+i\phi} \, .
\eeqa
Note that in our notations $\<Z|Z\>=\<\bar Z|\bar Z\>=1$, $\<\bar Z|Z\>=0$ and the spins  in the kets and bras are ordered from left to right, i.e. $\< ZX|ZX\>=1$ while $\< XZ|ZX\>=0$.
Summarizing, the difference between the two maps is most clearly illustrated by their action on the local basis
\beqa
\dagger &:&\quad \psi(n_1,\dots,n_N)\,| n_1,\dots,n_N\>\quad \mapsto\quad \psi^\dagger(n_1,\dots,n_N)\,\<n_1,\dots,n_N| \, ,\nn  \\
 \mathcal{F}  &:&\quad \psi(n_1,\dots,n_N)\,| n_1,\dots,n_N\>\quad \mapsto\quad \psi(n_1,\dots,n_N)\,\<L-n_N+1,\dots,L-n_1+1|\hat C \, , \nn
\eeqa
where $\hat C$ stands for charge conjugation, which exchanges $Z\leftrightarrow\bar Z$ and $X \leftrightarrow\bar X$.

We can now flip any Bethe state. Since the Hamiltonian is invariant under flipping the orientation of the chain and the charges, the operation $\cal F$ maps a ket Bethe state into another bra Bethe state. The overall factor depends on the normalization. To determine that overall factor in the coordinate normalization, consider first a two-magnon state
\beq
|\{u_1,u_2\}\>^{\co}= \sum_{1 \le n_1< n_2\le L} \[e^{ i p_1 n_1+ip_2 n_2}+S(p_2,p_1)  e^{  i p_2 n_1+i p_1 n_2} \] |n_1 , n_2\> \, .
\nn
\eeq
By changing variables to $m_1\equiv L-n_2+1$ and $m_2\equiv L-n_1+1$ it is easy to see that
\beqa
&&\mathcal{F}\circ|\{u_1,u_2\}\>^{\co}=
e^{i(L+1)( p_1+ p_2)} S(p_2,p_1)\,^\co\<\{u_1^*,u_2^*\}|\hat C  \,.\nn
\eeqa
This construction is trivially generalized to any number of magnons. We find
\beq
{\cal F}\circ |\{u_i\}\>^\co =e^{\{u\}}_L \, {g^{\{u-\frac{i}{2}\}}f_>^{\{u\}\{u\}}\over g^{\{u+\frac{i}{2}\}}f_<^{\{u\}\{u\}}}
\;\;\!^\co\<\{u_i^*\}|\hat C \,. \nn
\eeq
Our goal was to flip the second ket in (\ref{BreakBethe}) which we can now do. We have
\beq\la{flippedcor}
{\cal F}_{\text{second chain}}\circ |\{u_i\}\>^\co=\sum_{\alpha\cup\bar\alpha\in\{u\}} e^{\bar\alpha}_L  \, {g^{\bar\alpha-\frac{i}{2}}\over g^{\bar\alpha+\frac{i}{2}}}{
f^{\alpha\bar\alpha}f_>^{\bar\alpha\bar\alpha}f_<^{\alpha\alpha}\over f_<^{\{u\}\{u\}}}\ |\alpha\>_l^\co\otimes\,^\co\!\!_r\<\bar\alpha^*|\hat C \, .
\eeq
where $\bar \alpha^*$ is the set $\bar\alpha$ with its elements complex conjugated.
Similar expressions can be written in the algebraic conventions using the conversion factors derived before.
\footnote{Using (\ref{conversion}) we see that the algebraic Bethe states transforms as
\beq
{\cal F} \circ |\{u\}\>^\al=(-1)^N \ ^\al\<\{u^*\}|\hat C \, .
\eeq
Furthermore
\beq
{\cal F}_{\text{second chain}} \circ |\{u\}\>^\al=\sum_{\alpha\cup\bar\alpha\in\{u\}} f^{\alpha\bar\alpha}\, d_r^\alpha a_l^{\bar\alpha}\ \B^{\alpha}|0\>_l\otimes\!_r\<0|{\cal C}^{\bar\alpha^*}\hat C\ .
\eeq
Recall that in our normalization $\mathcal{C}(u^*)=-\left[ \mathcal{B}(u)\right]^\dagger$.}

\subsection{... and Sewing} \la{sewing:sec}
The last building block which we need to understand is the overlap of Bethe wave functions, i.e. scalar products. The procedure of overlapping the wave functions is what we denote as \textit{sewing}. The importance of spin chain scalar products in $\mathcal{N}=4$ SYM was pointed out in \cite{Roiban:2004va}.

The quantity we are interested in is\footnote{It is trivially related to the scalar product through $S^{\al}_N(\{v^*\},\{u\})=(-1)^N\,^\al\<\{v\}|\{u\}\>^\al$ which in turn follows from $\mathcal{C}(u^*)=-\left[ \mathcal{B}(u)\right]^\dagger$.}
\beq
S_N(\{v\},\{u\}) \equiv  \<0 | \prod_{j=1}^N \mathcal C(v_j) \prod_{j=1}^N \mathcal B(u_j) | 0 \> \, . \la{theguy}
\eeq
The explicit form of this quantity is actually known \cite{Korepin}. It is written for completeness in \eqref{genscalar} in appendix \ref{appscalar}.  If $\{u\},\{v\}$ take particular values this scalar product simplifies dramatically (luckily!). For example, when we consider the norm of a Bethe eigenstate we have $\{u\}=\{v\}$ and the Bethe roots obey the Bethe equations. In this case we have the remarkably simple result \cite{Gaudin}
\beq
S_N(\{u\},\{u\}) =  d^{\{u\}} \, a^{\{u\}}f_{>}^{\{u\}\{u\}}f_{<}^{\{u\}\{u\}}\det_{j,k} \partial_{j} \phi_k
\eeq
where $\phi_k$ was introduced in (\ref{BAEu}), $\partial_j=\frac{\partial}{\partial u_j}$ and we are using the shorthand notation introduced in \eqref{prodnot}. We shall also use $\det\limits_{j,k} \partial_{j} \phi_k \equiv \det \phi'_{\{u\}}$.
In the coordinate basis this formula reads
\beq\la{normcoord}
\N^{\,\co}(\{u\}) \equiv \ ^{\co}\<\{u\}|\{u\}\>^{\co}=
{1 \over g^{\{u+\frac{i}{2}\}}g^{\{u-\frac{i}{2}\}} }\, \frac{f_>^{\{u\;\}\{u\;\}}}{f_>^{\{u^*\}\{u^*\}}}\, \det \phi'_{\{u\}} \,.
 \eeq
As explained in section \ref{section2}, descendants of Bethe states are obtained by sending some of the roots to infinity (\ref{descendant}). We use $N$ to denote the total number of roots and $M$ to denote the number of finite roots.
The corresponding norm of a Bethe eigenstate descendant is related to (\ref{normcoord}) by the square of the Clebsch-Gordan coefficients
\beq\la{BetheNormDes}
\,^{\co}\<\{u;\infty^{N-M}\}|\{u;\infty^{N-M}\} \>^{\co}={(L-2M)! \, (N-M)!\over(L-M-N)!}\ ^{\co}\<\{u\}|\{u\} \>^{\co}\ .
\eeq

Another notable simplification occurs when we send one set of Bethe roots, e.g.\ the $v_j$, to infinity. In this case we find\footnote{Note that in  this case the algebraic Bethe states are not well normalized. Indeed, as we saw before,
$\mathcal C(u_j) \sim \displaystyle i \, a(u_j) / u_j \,S^+ \to \infty$ as $u_j\to \infty$. The coordinate wave functions are better normalized and do not vanish or diverge in this limit (\ref{descendant}).}
\beq\la{scalarBPS}
\,^\co\<\{\infty\}|\{u\}\>^\co={\,(-1)^N \, N!\over  g^{\{u+\frac{i}{2}\}}f_<^{\{u\}\{u\}}}\sum_{\alpha\cup\bar\alpha=\{u\}}(-1)^{|\alpha|} \,
e^{\alpha} \, f^{\bar\alpha\alpha}~,
\eeq
where the sum runs over all possible partitions of the $N$ elements of the set $\{u\}$ into subsets $\a, \bar \a$ and $|\a|$ denotes the number of elements in $\a$. Similarly, the inner product of a Bethe state descendant with a vacuum descendant is related to (\ref{scalarBPS}) by a product of Clebsch-Gordan coefficients
\beq
\,^\co\<\{\infty\}|\{u,\infty^{N-M}\}\>^\co = \frac{(L-M)! \, N!}{M! \, (L-N)!} \, \,^\co\<\{\infty\}|\{u\}\>^\co\ . \la{other2}
\eeq
In this formula the number of roots $u$ is $M$.

There is one more case where the general formula (\ref{theguy}) simplifies. This is the case when one of the operators is a Bethe eigenstate while the other one is left generic. This is presented in \eqref{GenBethe} in appendix \ref{appscalar}.

Note that sewing Bethe states involves at most sums over partitions of the magnons. The number of terms therefore only grows as a power of the number of magnons and is independent of the chains' lengths.

\subsubsection*{A new recursion relation for scalar products} \la{newsec}
For completeness let us illustrate how we could compute easily (\ref{theguy}) for arbitrary $\{u\}, \{v\}$. One possible strategy for computing this quantity is to derive a recursion relation yielding $S_N$ in terms of $S_{N-1}$. Such recursion relations exist in the literature, see e.g.\ \cite{Korepin}, and are presented in appendix \ref{appscalar} for completeness.

Here we derive a new recursion relation which we found quite efficient computationally. The idea is to get rid of particle creation operator $\mathcal B(u_1)$ by writing
\beq
S_N(\{v\},\{u\}) =  \<0 | \[\prod_{j=1}^N \mathcal C(v_j),\mathcal B(u_1)\]\prod_{j=2}^N \mathcal B(u_j)  | 0 \>
\eeq
which holds since the $\mathcal B$ operator kills the bra vacuum. Next we use the algebra relations (\ref{CB}) to compute this commutator. We have
\beq
\[\prod_{j=1}^N \mathcal C(v_j),\mathcal B(u_1)\]= \sum_n \(\prod_{j=1}^{n-1} \mathcal C(v_j) \) g(u_1-v_n) \(\mathcal A(v_n) \mathcal D(u_1) - \mathcal A(u_1) \mathcal D(v_n)\) \(\prod_{j=n+1}^{N} \mathcal C(v_j) \)\,. \la{commutator}
\eeq
The term between the two products is just $\[ \mathcal C(v_n) , \mathcal B(u_1) \] $. Now we pick the operators $\mathcal A$ and $\mathcal D$ in this expression and carry them to the left through the $\mathcal C$'s using the algebra relations. Recall that $\mathcal A$ and $\mathcal D$ have a trivial action on the left vacuum. This procedure does not change the number of $\mathcal C$'s which is now equal to $N-1$. So, what are the possible final arguments of the remaining $N-1$ $\mathcal C$'s? When commuting an $\mathcal A$ through a $\mathcal C$ we have
\beq
\mathcal C(v)\mathcal A(u) = f(v-u) \mathcal A(u) \mathcal C(v) + g(u-v) \mathcal A(v) \mathcal C(u) \, , \la{CA}
\eeq
which means that the $\mathcal A$ can keep the same argument or swap arguments with the $\mathcal C$. The same is true for a $\mathcal D$ operator passing through a $\mathcal C$. It is clear that we will get  terms where the arguments of the $\mathcal C$'s are $N-1$ out of the $N$ original $v$'s.
There is however one more possibility which comes from the algebra terms which swap the argument of the operators: we can also end up with $N-2$ of the original $v$'s together with the root $u_1$. That is
\beqa \la{newrecal}
&&S_N \(\{v_1,\dots,v_N\},\{u_1,\dots,u_N\}\) = \sum_{n} b_n \, S_{N-1} (\{v_1,\dots, \hat v_n,\dots,v_N\},\{\hat u_1, u_2,\dots,u_N\})\nn  \\
&&- \sum_ {n<m}  c_{n,m} \, S_{N-1}(\{ u_1,v_1,\dots \hat v_{n},  \dots , \hat v_{m},\dots v_N \}, \{\hat u_1,u_2,\dots,u_N\}),
\eeqa
where the hat means that the corresponding root is omitted. Naturally $S_0=1$. We simply need to find $b_n$ and $c_{n,m}$! Let us derive $b_n$. We start by re-ordering the $\mathcal C$'s in a smart way and write (\ref{commutator}) as\footnote{Note that we can order them in any way we want since they commute, $\[\mathcal{C}(u),\mathcal{C}(v)\]=0$, see table \ref{FZalg}.},
\beq
\[\(\prod_{j\neq n}^N \mathcal C(v_j) \)\mathcal C(v_n) ,\mathcal B(u_1)\] \, . \la{this}
\eeq
We are interested in $b_n$. In other words we want to consider the contribution to the scalar product where $\mathcal C(v_n)$ and $\mathcal C(u_1)$ are not present. This means that we must get rid of $\mathcal C(v_n)$ in (\ref{this}) and therefore the only term which will contribute is
\beq
\(\prod_{j\neq n}^N \mathcal C(v_j) \)\[\mathcal C(v_n) ,\mathcal B(u_1)\]=\(\prod_{j\neq n}^N \mathcal C(v_j) \) g(u_1-v_n) \(\mathcal A(v_n) \mathcal D(u_1) - \mathcal A(u_1) \mathcal D(v_n)\) \, . \la{nice}
\eeq
Now the $\mathcal A$'s must travel to the left using (\ref{CA}). Furthermore, when using these relations we must always pick the first term in the right-hand side. This is the term where the $\mathcal A$ and $\mathcal C$ do \textit{not} swap their arguments. Otherwise we would end up with a $\mathcal C$ with an argument $v_n$ or $u_1$ and therefore this would not contribute to $b_n$. For example,
\beq
\langle 0 | \(\prod_{j\neq n}^N \mathcal C(v_j) \) \mathcal A(v_n)=a(v_n) \prod_{j\neq n}^Nf(v_j-v_n)\,\langle 0 |\(\prod_{j\neq n}^N \mathcal C(v_j) \) +\dots
\eeq
where $\dots$ stand for terms where one of the $\mathcal C$'s ended up with the argument $v_n$. We used the fact that the action of $\mathcal A(v_n)$ on the bra vacuum simply yields $a(v_n)$. The other $\mathcal A$ and $\mathcal D$ operators in (\ref{nice}) are treated in the same way. It should be clear that at the end of the day we get
\beqa
b_n = g(u_1-v_n) a(v_n) d(u_1) \prod_{j\neq n} f(u_1-v_j) f(v_j-v_n)  + \(u_1 \leftrightarrow v_n\) ,  \la{bn}
\eeqa
With a similar reasoning we could derive $c_{n,m}$. We would find
\beqa
c_{n,m} = {g(u_1-v_{n}) \, g(u_1-v_{m}) \, a(v_{m}) d(v_{n})}{f(v_{n}-v_{m})} \prod_{j\neq n,m}^N  f(v_{n} - v_j)  \, f(v_j - v_{m}) + \(n \leftrightarrow m\) . \la{cn}
\eeqa
Since we know how to convert between the coordinate and algebraic Bethe states we can easily write a recursion relation for scalar products in the coordinate Bethe ansatz normalization. This is presented in appendix \ref{appscalar}.

The recursion relation (\ref{newrecal}) with (\ref{bn}) and (\ref{cn}) provides a complete solution to any scalar product in a straightforward way. For example, it is straightforward to implement this recursion in \verb"Mathematica".  As mentioned above, there exists another recursion for the general scalar product \eqref{theguy} in the literature, which we present for completeness in \eqref{oldrecal} in appendix \ref{appscalar}.

It is interesting to compare the two recursions in the following table:

\begin{center}
    \begin{tabular}{ p{8cm} | p{8cm} }
{\bf New recursion (\ref{newrecal})} & {\bf Usual recursion (\ref{oldrecal}) }\\ \hline
Uses scalar products with less particles as fundamental building blocks. & Uses generalized objects as building blocks. I.e., scalar products in different theories with different $a$ and $d$ functions appear at every step of the recursion and less particles. \\ \hline
Derived by reducing the number of particles in a very explicit way. Notion of particle is fundamental. & Derived from the analytic properties of the result. The scalar product is a rational function which can be therefore reconstructed from its poles and zeros. The notion of reducing the number of particles is secondary. \\ \hline
Contains much less terms than a brute force computation of the scalar product but still more terms than the recursion (\ref{oldrecal}). &
Contains much less terms than a brute force computation of the scalar product and less terms than the recursion (\ref{newrecal}).
\end{tabular}
\end{center}

Given this comparison it is fun to draw an analogy with two very important recursion relations in $\mathcal{N}=4$ SYM which are used in studying scattering amplitudes:  the CSW recursion relations \cite{CSW} and the BCFW expansion \cite{BCFW}. A very similar table would be suitable for comparing these two recursions.
The CSW and BCFW recursions would be the analogue of (\ref{newrecal})  and (\ref{oldrecal}) respectively.

\section{Three-point functions} \la{four}

Having introduced the necessary tools in the previous section, we are now ready to put them together to compute the objects of interest of this paper:  structure constants $C_{123}$ of three single trace operators in $\N=4$ SYM at tree level. An alternative way or writing the structure constants in a scheme independent way is as
\beqa
C_{123}&\equiv&\<\O_1 (x_1) \O_2 (x_2) \O_3 (x_3) \> \sqrt{\prod_{i=1}^3 {\<\O_i(x_{i+1})\bar\O_i(x_{i+2})\>\over\<\O_i(x_i)\bar\O_i(x_{i+1})\>\ \<\O_i(x_{i})\bar\O_i(x_{i+2})\>}}
\eeqa
where the indices are identified modulo $3$. The right hand side is indeed a $x_i$ independent quantity which coincides with $C_{123}$ once we use (\ref{2pf}) and (\ref{3pt}).
We want to study these structure constants at tree level, i.e. we are after the first term  in the expansion \eqref{expansion}. We will focus on a particular subset of operators as we will now describe.

\subsection{Setup}

The operators we will consider are linear combinations of single trace operators made out of two complex scalars. As reviewed in the previous sections, such operators with definite one-loop anomalous dimensions can be represented by Bethe eigenstates on a spin ${1\over2}$ chain.  This fact will allows us to use the technology introduced in the previous section for cutting and sewing the operators.

\begin{figure}[t]
\centering
\def\svgwidth{15.5cm}

\ifpdf
    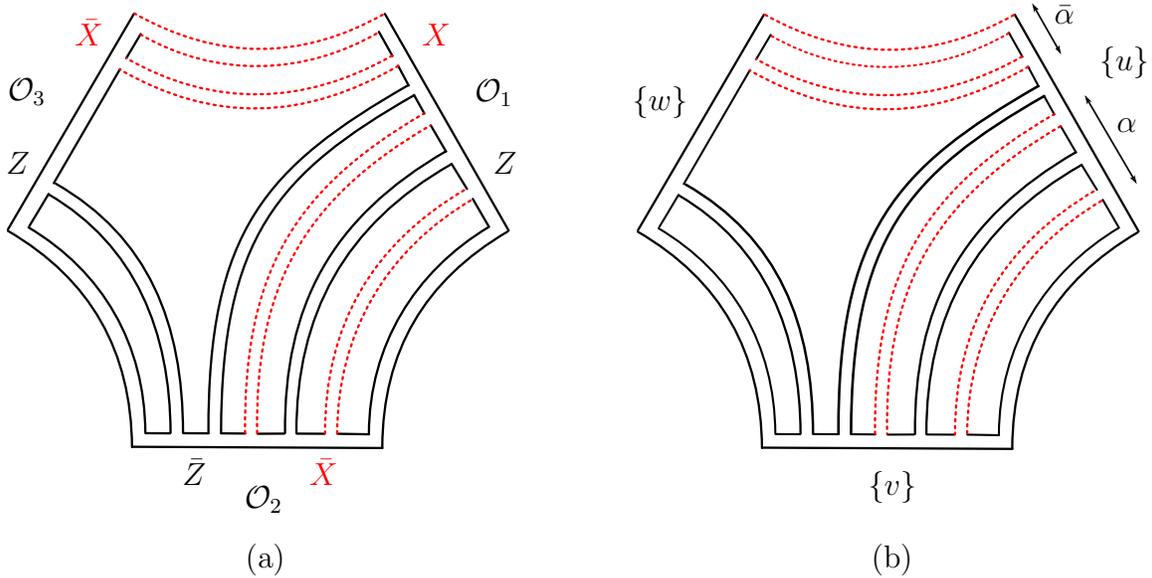
\else
    \com{PDF-picture replacement}
\fi

\caption{(a) Three-point function of $SU(2)$ operators at tree level.  All contractions are such that R-charge is preserved.  This is the simplest non-trivial configuration which is not extremal.  Note that the number of excitations on each chain is subject to the condition $N_1= N_2+N_3$. Also, if we denote by $l_{ij}$ the number of propagators between operators $i$ and $j$, we have $l_{12}=L_1-N_3$, $l_{13}=N_3$ and $l_{23}=L_3-N_3$. (b) We show the partitions of the excitations in $\O_1$.  Note that we only need to use the operation of cutting for this operator, as a part of its excitations $(\a)$ are contracted with those of $\O_2$ and another part $(\bar \a)$ with those of $\O_3$.}\label{3ptfunction}
\end{figure}

The general setup that we will use in the rest of the paper is presented in figure \ref{3ptfunction}a.   We consider three operators $\O_1$, $\O_2$ and $\O_3$,  with corresponding lengths $L_1$, $L_2$ and $L_3$. The number of propagators between operators $i$ and $j$ is fixed to be
\beq\la{lij}
l_{ij}= \frac{1}{2} (L_i + L_j -L_k),
\eeq
where $(i,j,k)$ is a permutation of $(1,2,3)$. We will restrict ourselves to the non-extremal case, that is where all $l_{ij}$'s are strictly positive. This is not only the generic case but it also has one important advantage: for non-extremal correlators the contribution of the operator mixing with double trace operators is suppressed in $1/N_c$ and need not be considered. Furthermore, there is considerable evidence in the literature that the extremal correlators can be considered as an analytic continuation of the non-extremal ones when some $l_{ij}\to 0$ \cite{Rastelli}.

Each operator $\O_i$ is a single trace operator made out of two complex scalars and is mapped to a spin chain state. One of the scalars we interpret as being the vacuum (or spin up) while the other scalar we interpret as excitations on that vacuum (or spin down). The total length of the operator is $L_i$ and the number of excitations is denoted by $N_i$. The different scalars for the different operators are summarized in the following table:\footnote{Note that there is no physical difference between $\O_1$ and $\O_2$. Indeed, our results will be invariant under the exchange of the two. The choice of vacuum (\ref{Vacchoice}) however is not symmetric. Still, we found this choice more convenient to work with than a more symmetric one.}
\beq\la{Vacchoice}
\begin{array}{ccc}
& \text{vacuum} & \text{excitations} \\ \hline
\O_1 & Z & X \\
\O_2 & \bar Z & \bar X \\
\O_3 & Z & \bar X
\end{array}
\eeq
Note that this is the only setup that is fully contained in the $SU(2)$ sector and which involves non-extremal correlators at the same time.
When Wick contracting the single trace operators with each other we can only connect a scalar with its conjugate. Therefore we see that all the vacuum constituents of $\O_3$ are connected $\O_2$ while all excitations of $\O_3$ are contracted with $\O_1$, i.e.
\beq
l_{23}=L_3-N_3 \qquad , \qquad l_{13}=N_3 \,.
\eeq
The richest contraction is between operators $\O_1$ and $\O_2$ at length
\beq
l_{12}=L_1-N_3\ .
\eeq
In this case we can overlap both vacuum and excitations. Finally, note that the total number of excitations in the three operators are constrained by
\beq
N_1 = N_2 + N_3 \, .
\eeq
These statements are illustrated in figure  \ref{3ptfunction}a.

\subsection{Brute force computation}

Representing each operator by a spin chain state of the form \eqref{psiCBA}, we can always compute the three-point function of figure \ref{3ptfunction}a by brute force as
\beq \la{3ptbf}
c^{(0)}_{123} =\alpha
\displaystyle \sum_{1\le n_1 < \dots <n_{N_2}\leq L_1-N_3} \psi^{(1)}_{n_1,\dots n_{N_2} , L_1-N_3+1,L_1-N_3+2,\dots , L_1} \, \psi^{(2)}_{L_2+1-n_{N_2} , \dots, L_2+1-n_{1}}\,\psi^{(3)}_{1,2,\dots,N_3}
\eeq
where $$\alpha= \sqrt{L_1\,L_2\,L_3\over\mathcal{N}_1\,\mathcal{N}_2\,\mathcal{N}_3}
$$ and $\psi^{(j)}$ is the wave function of $\mathcal{O}_j$ given in \eqref{waveN}.  The factor $\alpha$ contains some symmetry factors and the normalization of the states  $\N_j$ which is given by
\beq
\mathcal{N}_j= \sum_{1\le n_1 < \dots < n_{N_j} \le L_j} \(\psi^{(j)}_{n_1,\dots,n_{N_j}}\)^* \(\psi^{(j)}_{n_1,\dots,n_{N_j}}\) \, .
\la{normbf}
\eeq
This way of computing the structure constant is equivalent to the direct Feynman diagram computation and involves the precise form of the $N$-particle wave function (\ref{waveN}), which has $N!$ plane wave terms, and a sum of $O\((L_1-N_3)^{N_2}\)$ terms. For large number of excitations and long chains, \eqref{3ptbf} is very inefficient. This is when the techniques we introduced in the previous section will prove most powerful.  Namely, we will see that using the operations of \textit{cutting}, \textit{flipping} and  \textit{sewing}, one can express the structure constant purely in terms of the rapidities and lengths of each spin chain.

Note that the \textit{absolute value} of the structure constants is independent of the normalization of the three operators since we divide the overlap by the normalization of the wave functions. On the other hand changing the phase of these wave functions does not change their norm but it \textit{will} modify the overlap by a phase. Therefore the \textit{phase} of the structure constants is sensitive to the normalization of the operators. The phase is important however for bootstrapping higher-point functions and therefore we will keep track of it.\footnote{An analogy is the metric on curved spacetime which depends on the choice of coordinate system.} To fix this phase we chose to work with the coordinate convention for the wave functions.\footnote{Recall that the coordinate normalization is sensitive to the order of Bethe roots. For different orders we get therefore different structure constants. Of course, the absolute value of $C_{123}$ is always the same but the phase does change. }

\subsection{Tailoring three-point functions}
To compute the three-point function using the tools introduced in section \ref{sectailoring}, we first need to use the \textit{cutting} and \textit{flipping} operations to cut the three operators and prepare them to be \textit{sewed}. In principle each operator should be decomposed and flipped as in (\ref{flippedcor}). For the operator $\O_i$ the lengths of the left and right subchains in this formula are $l_{i-1,i}$ and $l_{i,i+1}$ respectively (\ref{lij}), (where the indices are identified modulo $3$). Recall that in this formula $\alpha$ is the subset of magnons on the left subchain and $\bar \alpha$ is the subset of magnons on the right subchain. Now, as depicted in figure \ref{3ptfunction}, only very particular subsets contribute when we cut the operators $\mathcal{O}_3$ and $\mathcal{O}_2$. Namely we need $\bar \alpha_3= \emptyset $ for $\mathcal{O}_3$ and $\alpha_2 = \emptyset$ for $\O_2$. The non-trivial cutting is that of $\O_1$. The length of the left and right spin chains in this case are $L_1-N_3$ and $N_3$ respectively.

The second step is to compute the scalar products between the different subchains, this is what we called the \textit{sewing} procedure above. Note that the contractions between operators $\O_2$ and $\O_3$ are trivial, as we are simply contracting vacuum fields.  On the other hand, the contractions between $\O_1$ and $\O_2$ and between $\O_1$ and $\O_3$ are nontrivial and we need to use the inner products of Bethe states to compute them. Finally, we normalize our result dividing it by the norm $\N_j$ of each $\O_j$. We use the notation shown in figure \ref{3ptfunction}b and denote the rapidities of $\O_1,\O_2,\O_3$ by $u,v,w$ respectively.

At the end of the day, the three-point function is given by
\beqa\la{structure}
c^{(0)}_{123}&=&
\sqrt{L_1\,L_2\,L_3\over\mathcal{N}_1\,\mathcal{N}_2\,\mathcal{N}_3}\,{e_{L_2}^{\{v\}}f_>^{\{v\}\{v\}}\over f_<^{\{v\}\{v\}}f_<^{\{u\}\{u\}}}\sum_{\alpha\cup\bar\alpha=\{u\}}e^{\bar\alpha}_{L_1+1}
\,{f^{\alpha \bar\alpha}
f_>^{\bar\alpha \bar\alpha} f_<^{\alpha \alpha}}
 \<\{v^*\}|\alpha\> \<\bar\alpha^*|\{w\}\>
\eeqa
where
\beq
e_l(u)\equiv{a_l(u)\over d_l(u)}=\(\frac{u+\frac{i}{2}}{u-\frac{i}{2}}\)^l\ ,\qquad f(u)=1+\frac{i}{u} \, . \la{adn}
\eeq
In these formulae all quantities are computed in the coordinate normalization as mentioned above.\footnote{As explained above, different normalizations of the wave functions yield the same absolute value for $C_{123}$ but different phase for the structure constant. In other words, we get the same absolute value for $C_{123}$ for the coordinate wave function with any order of Bethe roots or even for states in the algebraic normalization. But the phase will differ for all these cases. } We chose to omit the explicit ``co" superscript to make the formula less cluttered. So,
\beqa
\<\bar\alpha^*|\{w\}\> \equiv\!\!\!\!\!\!\!\!\!\! \ _{\quad\ \  N_3}^{\quad \quad \co}\<\bar\alpha^*|\{w\}\>^\co_{N_3}\  , \qquad \<\{v^*\}|\alpha\> \equiv\!\!\!\! \ _{\ L_1-N_3}^{\quad \quad \co}\<\{v^*\}|\alpha\>^\co_{L_1-N_3} \,.
\eeqa
and
\beq
\mathcal{N}_1=\;_{L_1}^\co\!\<u|u\>_{L_1}^\co
\ ,\qquad \mathcal{N}_2=\;_{L_2}^\co\!\<v|v\>_{L_2}^\co
\ ,\qquad
\mathcal{N}_3=\;_{L_3}^\co\!\<w|w\>_{L_3}^\co
\eeq
Our expression (\ref{structure}) is completely given in terms of inner products between Bethe states in the coordinate normalization. These inner products can be found using the recursion relation derived in the previous section
 or using the general formula (\ref{innercoor}). Thus we solved the problem of computing $c^{(0)}_{123}$ for generic states in our setup.

Furthermore, most inner products in (\ref{structure}) are not the most generic ones and therefore the formula can be simplified considerably. For example, for the normalization factors $\mathcal{N}_j$ we can simply use (\ref{normcoord}) which is of course much simpler than the generic inner product! The inner product $\<\bar\alpha^*|\{w\}\> $ is also quite simple: since the states $|\{w\}\> $ and $|\bar\alpha^*\> $ are proportional the vacuum descandant with all spins down (see figure \ref{3ptfunction}b), the product $\<\bar\alpha^*|\{w\}\>$ factorizes into\footnote{Again we omit the subscript $N_3$ and the superscript ``co" which is common to all ket's and bra's  in this expression.}
\beq
\<\bar\alpha^*|\{w\}\>={ \<\bar\alpha^*|\{\infty\}^{N_3}\> \<\{\infty\}^{N_3}|\{w\}\> \over  \<\{\infty\}^{N_3}|\{\infty\}^{N_3}\> }\ .
\eeq
For the inner products in this expression we can now use (\ref{scalarBPS}) and (\ref{other2}) which are again much simpler than the general case (\ref{innercoor}). In sum, the only complicated inner product is  $\<\{v\}|\alpha\>$. For this case we should indeed use (\ref{innercoor}) generically. There are some cases of course when even this inner product simplifies. For example, if we send the roots  $v$ to infinity or if we send the roots $u$ (and thus $\alpha$) to infinity. In other words, when either $\O_1$ or $\O_2$ is a BPS operator.

\subsection{General cases}

\begin{table}[h]
\beqa \nn
\begin{array}{lll}
\displaystyle C^{\circ\circ\circ}_{123}=
{\displaystyle {l_{12} \choose N_2}}
\frac{1}{{\caB}_1{\caB}_2{\caB}_3}  &,& \displaystyle
C^{\circ\bullet\circ}_{123}
=\(\!\!
\bea{c}
l_{12}-M_2\\
N_2-M_2
\eea\!\!
\){
{\caA}(l_{12}|\{v\})
\over {\caB}_1{\caB}_2{\caB}_3} \vspace{.3cm}\\  \displaystyle
C^{\circ\circ\bullet}_{123}
=
{\displaystyle {l_{12} \choose N_2}}
{
{\caA}(l_{13}|\{w\})
\over {\caB}_1{\caB}_2{\caB}_3} &, & \displaystyle C^{\bullet\circ\circ}_{123}
=\(\!\!
\bea{c}
l_{12}-M_1\\
N_2
\eea\!\!
\){
{\caA}(l_{13}|\{u\})
\over {\caB}_1{\caB}_2{\caB}_3} \vspace{.3cm}\\
\displaystyle C^{\circ\bullet\bullet}_{123}=
\(\!\!
\bea{c}
l_{12}-M_2\\
N_2-M_2
\eea\!\!
\)
{{\caA}(l_{12}|\{v\}){\caA}(l_{13}|\{w\})\over
 {\caB}_1{\caB}_2{\caB}_3 }&,& \displaystyle  C^{\bullet\circ\bullet}_{123}=
\(\!\!
\bea{c}
l_{12}-M_1\\
N_2
\eea\!\!
\)
{{\caA}(l_{13}|\{u\}){\caA}(l_{13}|\{w\})\over
 {\caB}_1{\caB}_2{\caB}_3 }   \vspace{-.15cm}
\end{array}
\eeqa
\beqa
\!\!\!\!\!\!\! && \!\!\!\!\!\!\!  \!\!\!\!\!\!\!   \!\!\!\!\!\!\! \!\!\!\!\!\!\!  \!\!\!\!\!\!\!\!\!\!    \!\!\!\!\!\!\!   \!\!\!\!\!\!\!  \!\!\! C^{\bullet\bullet\circ}_{123}={ 1 \over {\caB}_1{\caB}_2{\caB}_3}\sum\limits_{\scriptsize\begin{array}{c}\alpha\cup\bar\alpha=\{u\}\\|\bar \alpha|={l_{13}}\end{array}}
e_{L_{1}}^{\bar \alpha} f^{\alpha \bar \alpha}
{\caA}(l_{13}|\bar \a){\caC}(l_{12}|\a,\{v\}) \nn   \vspace{-.1cm} \\
\!\!\!\!\!\!\! && \!\!\!\!\!\!\!  \!\!\!\!\!\!\!   \!\!\!\!\!\!\! \!\!\!\!\!\!\!  \!\!\!\!\!\!\!\!\!\!   \!\!\!\!\!\!\!   \!\!\!\!\!\!\! \!\!\! C^{\bullet\bullet\bullet}_{123}={ {\caA}(l_{13}|\{w\}) \over {\caB}_1{\caB}_2{\caB}_3}\sum_{\scriptsize\begin{array}{c}\alpha\cup\bar\alpha=\{u\}\\|\bar \alpha|={l_{13}}\end{array}}
e_{L_{1}}^{\bar \alpha} f^{\alpha \bar \alpha}
{\caA}(l_{13}|\bar \a){\caC}(l_{12}|\a,\{v\}) \nn   \vspace{-.1cm} \nn
\eeqa\vspace{-.4cm}

\caption{Tree level structure constants $c^{(0)}_{123}$ for three $SU(2)$ operators in the setup of figure \ref{3ptfunction}. We use $\circ$ to indicate a BPS state and $\bullet$ to label a non-BPS state. So, for example, $C^{\circ\bullet\bullet}_{123}$ corresponds to the structure constants when $\O_1$ is a protected BPS operator while $\O_2$ and $\O_3$ are generic Bethe eigenstates. The most general case is of course $C^{\bullet\bullet\bullet}_{123}$. The last two cases were computed for highest weights only, i.e. for $M_a=N_a$. It is straightforward to generalize them to the case comprising descendants but the formulas become more involved. The first two cases are discussed in greater detail in the next subsection.} \la{finalresults}
\end{table}

In this section we summarize the final results for generic operators $\O_i$. To present the results in a concise way we should introduce some useful notation. As before we use $L_i$ to denote the length of operator $\O_i$, $N_i$ to denote the number of excitations of this operator and $M_i$ to denote the number of finite Bethe roots. The sets of finite roots of the three operators are denoted by $\{u\},\{v\},\{w\}$.

We further introduce
\beqa
{\caB}_a&\equiv &
\frac{g^{\{v-\frac{i}{2}\}}f_<^{\{v\}\{v\}} }{\sqrt{g^{\{v+\frac{i}{2}\}}g^{\{v-\frac{i}{2}\}}}}\sqrt{\frac{f_>^{\{v\;\}\{v\;\}}}{f_>^{\{v^*\!\}\{v^*\!\}}}\frac{\det\phi'_{\{v\}}}{L_a}
\(
\bea{c}
L_a-2M_a\\
N_a-M_a
\eea
\)} \la{Beq}
\\
{\caA}(l|\{ u\})&\equiv &
\sum\limits_{\alpha \cup \bar\alpha =\{u\}}(-1)^{|\alpha|}f^{\alpha\bar\alpha}/e_{l}^{\alpha}
\eeqa
where $\phi_j$ is introduced in (\ref{BAEu}),  $a_l,d_l, f$ are defined in (\ref{adn}) and $\alpha$ and $\bar \alpha$ are partitions of $\{u\}$. As usual, we are using the shorthand notation (\ref{prodnot}) and (\ref{morenotation}). The set of Bethe roots appearing in ${\caB}_a$ is $\{u\},\{v\},\{w\}$ for $a=1,2,3$ respectively; furthermore recall that these sets comprise only the $M_a$ finite roots of each operator.
Finally we denote
\beqa\la{calC}\nn
{\caC}(l|\{ u\},\{ v\})&\equiv &\sum_{\scriptsize\begin{array}{c} \beta \cup \bar\beta = {\{ u\}} \\
\a \cup \bar\a={\{v\}}  \end{array}}g_<^{\{u\}\{u\}}g_>^{\{v\}\{v\}} (-1)^{P_{\beta}+P_{\a}}
 {e_{l}^{\bar\beta}e_{l}^{\alpha}}  h^{\beta\alpha} h^{\bar\alpha\bar\beta} h^{\beta\bar\beta} h^{\bar\alpha\alpha}
\det t^{\beta\alpha}\det t^{\bar\a\bar\beta}
 \;.
\eeqa
where $(-1)^{P_{\a}}$ is defined as a sign of permutation of the ordered set $\{v\}$ which gives $\alpha\cup\bar\alpha$\footnote{In {\it Mathematica}
we define $(-1)^{P_{\a}}$ as
$\rm sign[a\_List, ab\_List, v\_List] :=
 Signature[Join[a,ab]] Signature[v]$
}. Finally
\beq
h(u)=1-iu\ , \qquad t(u)=-\frac{1}{u(u+i)}
\eeq
In addition to our usual shorthand notation we are also using
\beq
\det t^{\beta\alpha} \equiv  \det_{\scriptsize \begin{array}{c} {u_i} \in \beta\\ v_j \in \alpha\end{array}} t(u_i-v_j)
\eeq
In terms of these useful functions the final results for $c_{123}^{(0)}$ are presented in table \ref{finalresults}. Note that $C^{\circ\bullet \circ}_{123}$ and $C^{\bullet \circ \circ}_{123}$ as well as between $C^{\circ\bullet\bullet}_{123}$ and $C^{\bullet \circ\bullet}_{123}$ are not different cases. That is, $\O_1$ and $\O_2$ are exchanged if we choose to view the same operators as $Z$'s ($\bar Z$'s) excitation of the $X$'s ($\bar X$'s) instead of $X$'s ($\bar X$'s) excitations of the $Z$'s ($\bar Z$'s). We notice that $C^{\circ\bullet\circ}_{123}$ and $C^{\circ\bullet\bullet}_{123}$ are only nonzero when $M_2\le l_{12}\le L_2-M_2$ due to cancelations in ${\caA}_2(l_{12})$.

We now move to the discussion of the two simplest cases which are the cases when all operators are BPS or when only $\O_2$ is not BPS.

\subsection{Simplest examples}
The simplest three-point functions involves three protected (BPS) operators. This case is protected by supersymmetry and was studied in \cite{Lee:1998bxa}. The tree-level result holds at any coupling. In our language this correlation function involves three operators where all excitations have zero momentum, i.e. all Bethe roots are sent to infinity. In other words, all three operators are vacuum descendants in the $SU(2)$ sense. In this case
(\ref{structure}) simplifies to a simple combinatorial factor\footnote{The normalization of the coordinate Bethe ansatz state when all roots are sent to infinity is given by $\ ^\co\< \{\infty\}^N |\{\infty\}^N\>^\co=\displaystyle{L \choose N}(N!)^2$, see (\ref{other2}).}
\beq
c^{(0)}_{123}\equiv C_{123}^{\circ \circ \circ}={\sqrt{L_1L_2L_3}{\displaystyle {L_1-N_3 \choose N_2}} \over{\sqrt{\displaystyle {L_1 \choose N_1} {L_2 \choose N_2} {L_3 \choose N_3} }}}\ .
\eeq

The next to the simplest case is when one of the operators is not protected. In this case the three-point function is no longer fixed by (super) symmetry and it has a very interesting and rich structure. Consider for example the case when $\O_1$ and $\O_3$ are BPS states (i.e. vacuum descendants) while $\O_2$ is a generic Bethe eigenstate. In this case, the structure constant $c_{123}^{(0)}$ will be a function of the $N_2$ Bethe roots characterizing the operator $\O_2$. If all these roots are finite the operator corresponds to an highest weight spin chain state. When $N_2-M_2$ of these roots are sent to infinity we obtain $SU(2)$ descendants. I.e. we generate the full $SU(2)$ multiplets by acting on the highest weights with $\(S^-\)^{N_2-M_2}$. The number of finite roots is $M_2$ and we use $\{v\}$ to denote the set of finite roots.
The structure constant $c^{(0)}_{123}\equiv C_{123}^{\circ \bullet \circ}$ for the case when only $\O_2$ is non-BPS is then given by
\beqa
 C_{123}^{\circ \bullet \circ}=
\frac{ \sqrt{L_1L_2L_3} {L_1-N_3-M_2 \choose N_2-M_2}}{\sqrt{{L_1 \choose N_1} {L_3 \choose N_3}  {L_2-2M_2 \choose N_2-M_2}}}  \sqrt{ g^{\{v+\frac{i}{2}\}} g^{\{v-\frac{i}{2}\}} f_>^{\{v^*\}\{v^*\}} \over f_>^{\{v\}\{v\}}\, \det \phi'_{\{v\}}}\,
{{  \sum\limits_{\beta\cup\bar\beta =\{v\}} (-1)^{|\beta|} \,
 f^{\beta\bar\beta}/e_{L_1-N_3}^\beta } \over g^{\{v+\frac{i}{2}\}}f_<^{\{v\}\{v\}}}\ . \la{lastresult}
\eeqa
There are several interesting limits which we can take in (\ref{lastresult}). We can consider many magnons, few magnons, the near BMN limit, the classical limit etc.

For example, the simplest possible nontrivial case is when $\O_2$ has only two particles with opposite momenta. In this case we have simply $M_2=N_2=2$ and $v_1=-v_2\equiv v \equiv \frac{1}{2}\cot\frac{p}{2}$.  Furthermore, from the Bethe equations \eqref{BAE} we can see that in this case $p=2 \pi n /(L_2-1)$, with $n \in \mathbb{Z}$. Then
\beq\la{Twomag}
 C_{123}^{\circ \bullet \circ}=e^{-ip/2}  \sqrt{\frac{L_1L_2L_3}{ 2{L_1 \choose N_1} {L_2\choose 2}  {L_3 \choose N_3}}} \,{\sin \(\frac{p\, l_{12}}{2}\) \sin \(\frac{p}{2}(l_{12}-1) \)\over \sin^2 \(\frac{p}{2}\) },
\eeq
where $l_{12}=(L_1+L_2-L_3)/2$, $N_3=(L_1+L_3-L_2)/2$ and $N_1=N_3+2$.\footnote{Note that (\ref{Twomag}) is invariant under $l_{12}\to l_{32}=(L_2+L_3-L_1)/2$ which is a particle hole symmetry.}

The BMN limit is discussed in appendix \ref{BMNsec}.
We briefly comment on the classical limit in the next subsection. A more thorough analysis, comprising in particular a thorough discussion of the classical limit of all our results, will be reported elsewhere \cite{toappear}.

\subsection{Scaling limit}
\begin{figure}[t]
\centering
\includegraphics[scale=0.4]{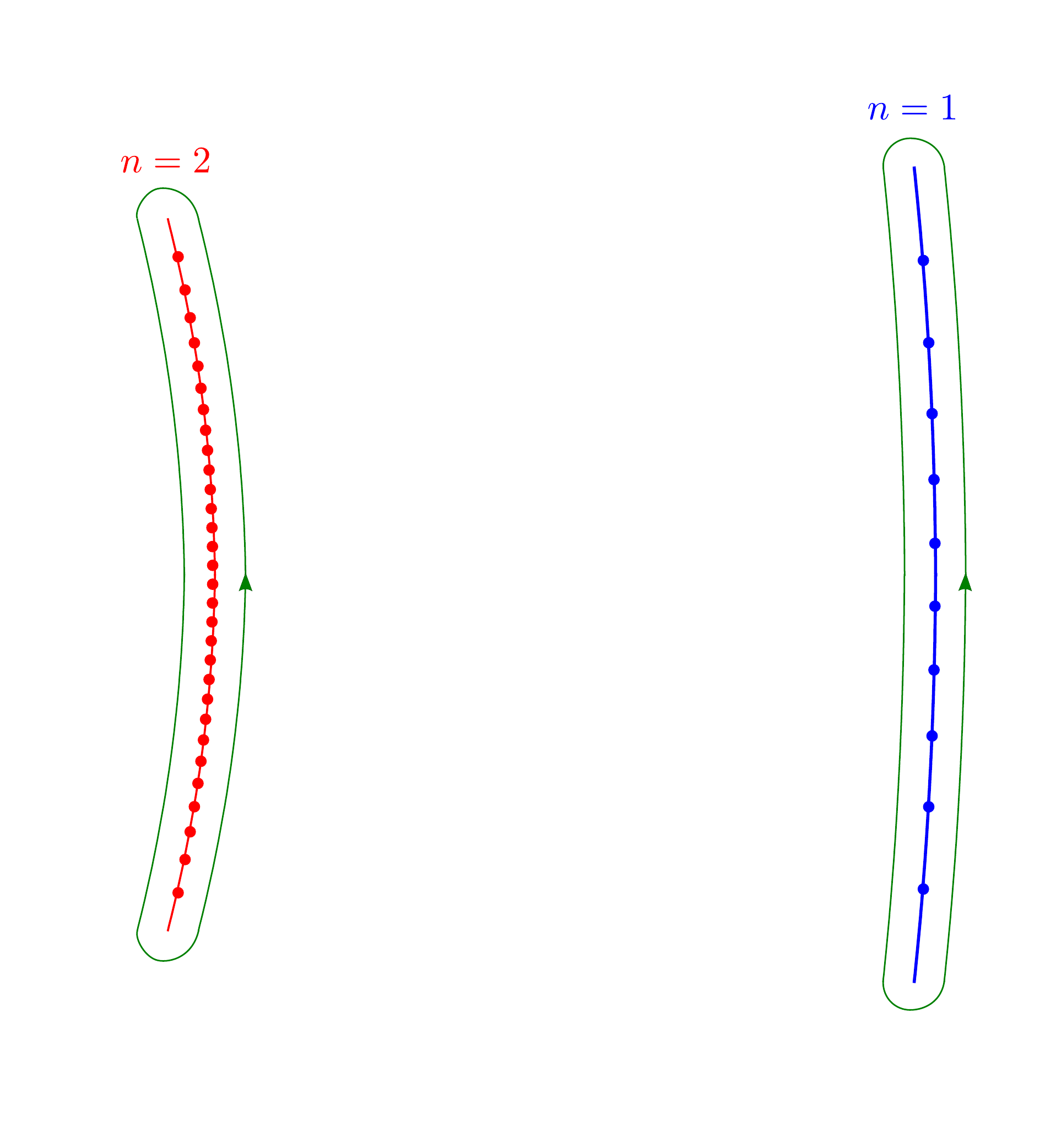}
\caption{In the classical limit the Bethe roots distribute themselves along disjoint cuts in the complex plane. The resolvent (\ref{res}) has square root cuts at the position of the Bethe roots in the continuum limit and defines a Riemann surface.  The final result for the structure constants (\ref{chapeau}) is given by some  integrals in this surface. The integral involving the quasi-momenta $q(v)$ is over the $A$-cycles of the surface denoted in green in the figure. The integral involving the density $\rho(u)$ is taken over the cuts. $n=1,2$ are the mode numbers in this real data example.}\label{cuts}
\end{figure}

The scaling (or thermodynamical) limit is also known as the Sutherland limit \cite{Sutherland:1995zz}
and corresponds to low lying excitations around the ferromagnetic vacuum.
In the AdS/CFT context this limit was re-discovered in \cite{Beisert:2003xu} and
proves to be a very useful limit for comparison with string theory computations through the so-called Frolov-Tseytlin limit \cite{Frolov:2002av}
which we describe below.
In the scaling limit the length $L$ and the number of Bethe roots $M$ is very large. The Bethe roots also scale to infinity as
\beq\la{ML}
M \sim L \sim v \to \infty
\eeq
and are of the same order. In this limit the roots distribute themselves in umbrella shaped contours $\mathcal{C}_k$ described by some density $\rho(v)$, see figure \ref{cuts}. These contours are also called \textit{cuts}. A systematic description of the spectrum  in the classical limit is achieved through the finite gap method of
KMMZ
\cite{Kazakov:2004qf}. In this method one introduces a resolvent
\beq
G(u)=\sum_{k=1}^{M}\frac{1}{u-v_k}=\int\limits_{\cup \,\mathcal{C}_k} dv \,\frac{\rho(v)}{u-v}
\la{res}
\eeq
where the integral is over all the contours $\mathcal{C}_k$ where the roots lie in the continuum limit.
The (derivative of) this function describes a hyperelliptic curve.
Our result can be written very neatly in this language. We find that the structure constants reduce to nice integrals on this Riemann surface, \cite{toappear}. For example, for the structure constants $C^{\circ\circ\bullet}$, $C^{\circ\bullet \circ}$ and $C^{\bullet\circ\circ}$ we find that the building blocks
\footnote{It may be interesting to understand the relation to the string  splitting and joining approach of \cite{Casteill:2007td}.}
\beq
{{{\caA}(l|\{v\})
\over {\caB}(\{v\})}}
\sim \, \exp \int\limits_0^1 dt \(\,\oint\limits_{\cup \,\mathcal{C}_k} \frac{du}{2\pi i}  \,q\log (1-e^{iqt})-   \int\limits_{\cup \,\mathcal{C}_k} du \,\rho\,\log\(2\sinh(\pi t \rho)\)\,\) \la{chapeau}
\eeq
(up to a phase factor) where\footnote{In our normalization $\rho(u),q(u)\sim 1$ whereas $\int\rho(u)du=M\sim L$. This implies that the argument of the exponent
scales as $L$.}
\beq
q(u)=\frac{l}{u}-G(u) \la{qdef}
\eeq
is a sort of trimmed quasi-momenta, (see figure \ref{cuts}). In this expression all the roots are taken to be finite. Expression (\ref{chapeau}) holds for any scaling configuration of roots $\{v\}$ with any number of cuts.\footnote{For some configurations with large filling fractions we might find some logarithmic singularities in the first integral. The safest way to compute it in this case is to introduce some
twists to regularize the integrals as done in \cite{Gromov:2007ky} and then analytically continue the results to the case of zero twist. } The scaling limit of the more general cases in table \ref{finalresults} can be also analyzed. A detailed analysis of the classical limit and other interesting regimes will be published elsewhere.

For concreteness let us consider the case where $\mathcal{O}_2$ is non BPS while $\mathcal{O}_1$ and $\mathcal{O}_3$ are protected.
The simplest example of the configuration with the scaling \eq{ML} is the configuration with one cut. Appart from the length $L_2$
and the number of roots $M_2\equiv \alpha L_2$ it has one more integer parameter - the mode number $n$
which fixes the $2\pi i n$ ambiguity of logarithm of the BAE (for more details see \cite{Kazakov:2004qf}).

This solution is dual to a string state which, at strong coupling is described by a classical string motion,  the rigid circular string.
For concreteness let us briefly describe this solution. The string is restricted to $R\times S^3\subset AdS_5\times S^5$, where $R$ is the global $AdS_5$ time $t$ and the $S^3\subset S^5$ can be describe by four embedding coordinates $X_i$
satisfying $\sum_i X_i^2=1$.
The circular string classical solution is given by
\beq
X_1+iX_2=\sqrt{\frac{{\cal J}(1-\alpha)}{w_1}}e^{iw_1\tau+i m\sigma}\;\;,\;\;
X_3+iX_4=\sqrt{\frac{{\cal J}\alpha}{w_2}}e^{iw_2\tau+i (m-n)\sigma}\;\;,\;\;t=\kappa\tau
\eeq
where $\sigma$ and $\tau$ are the worldsheet coordinates in conformal gauge. The map to the spin chain state is established by
$L_2=\sqrt\lambda{\cal J}$ and $M_2=\alpha\sqrt\lambda{\cal J}$. The others parameters of the solution are not all independent due to the equations of motion and the Virasoro
constraints.
The level matching condition gives $m=\alpha n$
whereas the other constraints are more complicated and in the
large ${\cal J}$ limit they give
\beq
\kappa\simeq
{\cal J}+\frac{\alpha(1-\alpha)n^2}{2{\cal J}}
\;\;,\;\;
w_1\simeq {\cal J}+\frac{\alpha(2\alpha-1)n^2}{2{\cal J}}\;\;,\;\;
w_2\simeq {\cal J}+\frac{(\alpha-1)(2\alpha-1)n^2}{2{\cal J}}\ .
\eeq
The classical string energy is $\Delta=\sqrt\lambda\kappa$.

Notice that ${\cal J}=L_2/\sqrt\lambda$ can be very large even for small couplings $\lambda\sim 0$. This implies that
the large $\cal J$ expansion of the classical string energy
might capture some information about the weak coupling expansion.
Indeed, expanding the anomalous dimensions in the scaling limit \eq{ML}
the one-loop spin chain gives
\beq
\gamma_2=\lambda\frac{\alpha(1-\alpha)n^2}{2L_2}+\O\(\frac{\lambda}{L_2^2}\)
\eeq
which is precisely $\Delta-L_2$. This is a particular example of the FT limit alluded above.

We hope that, similarly, the scaling limit of the structure constants \eq{chapeau} can be obtained from the classical string theory in a similar limit.

For illustration let us  use the general expression (\ref{chapeau}) to evaluate  $C_{123}^{\circ \bullet \circ}$ in the small filling fraction $\alpha$ expansion. For simplicity we will focus on the case when $M_2=N_2$, i.e. the circular string is a highest weight,
and expand for small $\alpha=M_2/L_2$.
It is useful to introduce
$\beta\equiv(L_1-N_3)/L_2$ and $r\equiv \pi n \beta$.
We find (see appendix \ref{AppQC} for more details)
\beq
\left| C_{123}^{\circ \bullet \circ} \right| = \left| C_{123}^{\circ \circ \circ} \right| \exp \[ L_2 \,\Gamma
+ \mathcal{O}\(\frac{1}{L^0_2}\) \]\la{circular}
\eeq
where
\beq
\Gamma =\alpha  \log \left(\frac{\sin r}{r}\right)+\frac{\alpha ^2}{2} \left[\frac{r^2}{3 \beta ^2}+\frac{(\beta -1)}{\beta}
  \left(  \frac{r^2}{\sin^2 r}-1\right) \right] + \mathcal{O}(\alpha^3) \,.
  \eeq

\section{Conclusions, speculations and open problems}\la{concl}
Computing three-point functions in $\mathcal{N}=4$ super Yang-Mills is an ambitious goal with far reaching consequences. Together with the two-point functions, the three-point functions are enough to reconstruct any higher-point function in this conformal field theory. It is fascinating to note that integrability techniques, so useful in the spectrum problem, can be used to tackle the computation of structure constants at weak coupling.

Basically, for the spectrum problem, we needed to know the spectrum of integrable Hamiltonians. For the structure constants we require the eigenvectors as well, i.e.\ the precise structure of the spin chain wave functions. Using integrability we computed all tree-level non-extremal three-point functions of $SU(2)$ sector operators, see table \ref{finalresults}. The method used involved cutting and gluing single trace operators in a very stringy operation, see figure \ref{3D3P}. It would be fascinating if we could interpret our results in a string field theoretic framework as advocated in \cite{Okuyama:2004bd}. The BMN limit might be a good starting point \cite{Spradlin:2002ar}.

There are many interesting directions to pursue.

We want to generalize our results to the full field strength multiplet of $\mathcal{N}=4$ SYM. After all, to glue three-point functions together into higher-point correlators we will need the most general cases. To complete this task we need to develop new technology for computing scalar products of Bethe states for higher rank groups such as $PSU(2,2|4)$. We will present a more detailed study of the Nested cases elsewhere \cite{toappear2}. Let us simply anticipate a few results.

As a first step we generalized the formula for the normalization of Bethe states to generic Lie (super) algebras $r$. We found a remarkably simple formula generalizing (\ref{normcoord}) which depends neatly on the Cartan matrix $M_{a,b}$ and on the Dynkin labels $V_a$ of the corresponding algebra and representation. In that case the relevant phases are given by\footnote{To our knowledge the only previous results for scalar products of Bethe states in Nested Bethe ansatz systems are \cite{su3} for the norm of Bethe eigenstates in the $SU(3)$ spin chain and \cite{hub} for the norm of Bethe eigenstates in the Hubbard model. We could not find any result for more generic inner products in the literature.}
\beq
\phi^{(j)}_a = \frac{1}{i} \log \[\(\frac{u^{(j)}_a + \frac{i}{2}V_j }{u^{(j)}_a - \frac{i}{2}V_j }\)^L \prod_{j'=1}^r \prod_{\substack{b=1 \\ (j,a) \neq (j',b) }}^{K_{j'}} \frac{u^{(j)}_a - u^{(j')}_b -\frac{i}{2}M_{j,j'}  }{u^{(j)}_a - u^{(j')}_b +\frac{i}{2}M_{j,j'}  } \],
\la{genphases}
\eeq
where $u_a^{(j)}$ are the several Bethe roots of the corresponding Nested Bethe ansatz and there are $K_j$ roots of each type where $j=1,\dots,r$. For example, for the fundamental representation we have $V_j=\delta_{j,1}$. The normalization of the coordinate Bethe wave functions, with a convention analogue to (\ref{waveN}) simply reads \cite{toappear2}
\beq
\N_\co= \frac{(-1)^{\sum\limits_{j=2}^r K_j}}{g^{\{u^{(1)}+i/2\}}g^{\{u^{(1)}-i/2\}}} \, \det\limits_{I,J}\partial_{I} \phi_J \, ,
\la{normconj}
\eeq
where we combined the indices $a$ and $j$ into a single index $I$.

For the other rank one sectors of $\mathcal{N}=4$, i.e. for the $SL(2)$ and the $SU(1,1)$ sectors, we went a bit further. For these cases we generalized all possible scalar products (and not only the norm); this is discussed in section \ref{sl2su11}. With these results we can trivially convert the results of the main text to their analogue in the other rank one sectors.\footnote{For example, we can go from the $SU(2)$ to the $SL(2)$ sector by formally replacing $L\to -L$ in the several formulae (as usual at weak coupling).}

Another possible direction to generalize to more generic operators might be to consider the superprotected correlators of \cite{Drukker:2009sf} as a vacuum and study excitations around these.

One should also move to higher loops. There are two type of loop corrections. One type are loop corrections to the contraction of the three operators. At low loop orders, these are local corrections which dress the nonlocal overlap of the three wave functions which is studied in this paper. At one loop, and in the $SO(6)$ sector,  they are captured by Hamiltonian density insertions at the spin chain breaking points \cite{Okuyama:2004bd,Roiban:2004va,Alday:2005nd}. The insertion measures the energy cost of splitting the three operators \cite{Okuyama:2004bd}.

The other type of corrections are loop corrections to the wave functions of the three operators. These comprise nonlocal corrections -- due to the dependence of the scattering matrices (\ref{smatrix}) on the coupling -- and also local corrections due to the insertion of the so-called fudge factors, or contact terms \cite{Staudacher:2004tk}.
Due to the degeneracy of the spectrum at tree level, the computation of these corrections to the structure constants at $l$ loops involves the $(l+1)$ loops wave functions.

In appendix \ref{dataAp} we list some basic examples of one-loop structure constants computed by taking into account these two types of corrections.

Both loop corrections mentioned above could of course be incorporated in our treatment in a systematic way. In the meantime, given the experience with the spectrum problem, we could also try to make some educated guess from the analytic expressions for the one-loop structure constants. In the $SU(2)$ spectrum problem, up to four loops, all one needed to modify in the asymptotic Bethe equations was the expression for the energy and the momentum of the excitations, e.g.
\beqa
&&e^{ip} = \frac{u+i/2}{u-i/2} \,\,\to\,\, \frac{x(u+i/2)}{x(u-i/2)} \la{momrep}
\eeqa
where
\beq{\sqrt{\lambda} \over 2\pi}\,x(u)={u+\sqrt{u^2-\frac{\lambda}{4\pi^2}}}
\eeq
is the so-called Zhukoswky variable, introduced in \cite{Beisert:2004hm}. If we look at our final expressions, e.g.\ at (\ref{lastresult}), we see that the building blocks are the same building blocks arising in the spectrum problem. It is therefore very tempting to try the simple replacement (\ref{momrep}) and check whether our expressions could hold to higher loop order.

By comparing the results obtained in this way with the brute force results of table \ref{tabledata} we conclude that such simple replacement does \textit{not} work.

On the other hand, given our discussion of the two type of loop corrections this is perhaps not surprising. Our treatment ignores contact terms and loop corrections to the Wick contractions. What we  expect is that the replacement (\ref{momrep}) captures the correct result for large (and maybe dilute) spin chains where the correction to the $S$-matrix is the dominant effect. In contradistinction,  the examples in the table \ref{tabledata} involved small operators. It could be that something a bit more sophisticated than the naive replacement (\ref{momrep})   works even for small operators. We are currently investigating both possibilities.

\begin{figure}[t]
\centering
\def\svgwidth{15cm}
\ifpdf
    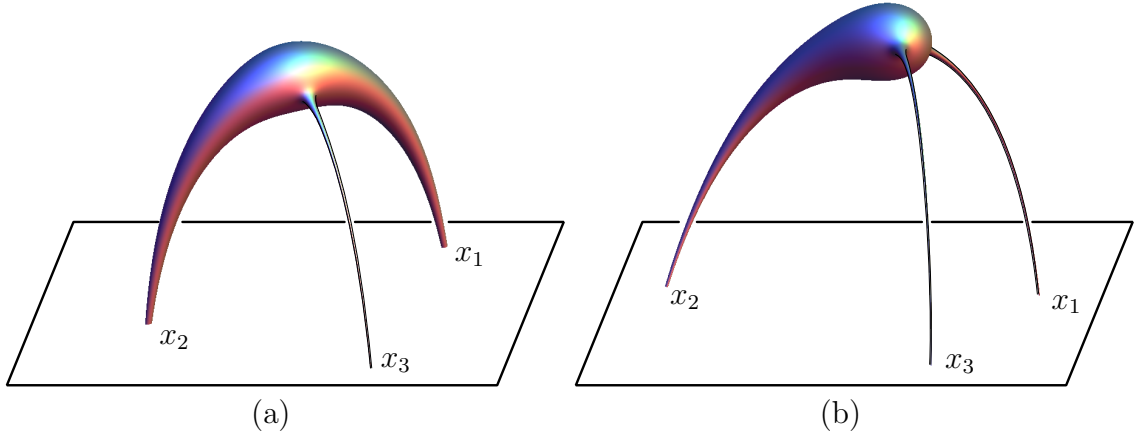
\else
    \com{PDF-picture replacement}
\fi

\caption{AdS classical string solutions describing a three-point function in the CFT. (a) A correlation function between two heavy operators and one light operator is depicted. For examples of such computations at strong coupling see \cite{recentpapers,recentpapers2}. (b) A decay of a heavy state into two light operators is depicted. Our prediction for the structure constant in this case is given by (\ref{lastresult}). It would be great if a simple upgrade of this expression \textit{a la} KMMZ would produce the correct string theory value. In particular, it would be very interesting to study this process directly at strong coupling and check whether it matches (\ref{lastresult}) in the FT limit.}\label{Insect1}
\end{figure}

We should also try to explore the classical limit of the structure constants to make contact with strong coupling computations using the dual string sigma model. In the Frolov-Tseytlin (FT) limit
\beq
\lambda , J  \to \infty \qquad \text{with} \qquad \lambda'\equiv \frac{{\lambda}}{J^2} \ll 1
\eeq
the strong coupling string energy admits an expansion in $\lambda'$ which resembles a weak coupling expansion \cite{FT}. It turns out that the first two loop orders of the weak coupling anomalous dimensions are reproduced by this expansion.\footnote{It is now understood that there is no deep reason for such match given the order or limits involved. Still it is a lucky coincidence that it works and we might also be fortunate to find it for the three-point functions.}  The match of weak and strong coupling results in the FT limit is most elegantly done through the algebraic curve method developed by KMMZ \cite{KMMZ}. In this work it was explained that any string classical motion can be mapped to a Riemann surface defined by some quasimomenta $\tilde q(x)$. The same is true at weak coupling: each solution to Bethe equations in the classical limit is described by some quasimomenta $q(u)$. The main difference between the weak and the strong coupling  quasi-momenta is in their pole singularities: in properly chosen variables the strong coupling quasimomenta has two poles at $x=\pm \lambda'$ while the weak coupling quasi-momenta has a single pole at $u=0$, see (\ref{qdef}). In the FT limit the strong coupling poles go to zero and the weak and strong coupling curves coincide. This establishes a precise match of the spectrum in the FT limit for any classical state \cite{KMMZ}.

Optimistically, the same might be true for three-point functions. It would be great if the strong coupling result in the FT limit reproduced our classical result (\ref{chapeau}) which holds for any classical state. It might be simpler, at strong coupling, to consider more restricted string states. In (\ref{circular}) we presented the outcome of our result for the structure constant involving the operator dual to the rigid circular string and two BPS states. Correlation functions of one heavy operator and two light operators were not yet studied at strong coupling, see figure \ref{Insect1}b. Needless to say, it would be extremely interesting to do so.

On the other hand, at strong coupling, there has been a lot of recent activity studying the correlation function of two heavy states and one light state \cite{recentpapers,recentpapers2}, see figure \ref{Insect1}a. These works followed related earlier works on two-point functions \cite{Janik:2010gc}. It would be very interesting to take the continuum limit of our expressions in that case and compare with the strong coupling results (in the FT limit). We will report on a much more detailed study of the classical limit of our results elsewhere \cite{toappear}.

\section*{Acknowledgments}

We thank J. Caetano, J. Gomis, R. Janik, J. Plefka, V. Kazakov, V. Korepin, C. Kristjansen, J. Maldacena, J. Penedones, L. Rastelli, F. Smirnov, A. Tseytlin, K. Zarembo and K. Zoubos for useful discussions. The research of J.E., A.S. and P.V. has been supported in part by the Province of Ontario through ERA grant ER 06-02-293. Research at the Perimeter Institute is supported in part by the Government of Canada through NSERC and by the Province of Ontario through MRI. P.V. would like to thank NBI and King's College for warm hospitality. This work was partially funded by the research grants PTDC/FIS/099293/2008 and CERN/FP/109306/2009.

\appendix

\section{Scalar products} \la{appscalar}

In this appendix, we write several useful formulas for the scalar product $S_N (\{v\}, \{u\})$ for $SU(2)$ spin chains in both the algebraic and coordinate bases.  These complement the formulas  in section \ref{sectailoring}. We also give explicit expressions for the scalar products for $SL(2)$ and $SU(1|1)$ spin chains. In order to simplify the expressions, it is convenient to introduce the following functions
\beq
h(u)={f(u)\over g(u)}={u+i\over i}~,\qquad t(u)={g^2(u)\over f(u)}={-1\over u(u+i)} \, ,
\eeq
and make use of the shorthand notation introduced in \eqref{prodnot}.

\subsection{Scalar products in $SU(2)$}

\subsubsection*{General scalar product as a sum over partitions}
If all rapidities involved in the scalar product are arbitrary complex numbers, the scalar product can be written as a sum over all possible partitions of the two sets of rapidities.  Explicitly \cite{Korepin, Slavnov}
\beq
S^{\al}_N(\{v\},\{u\})=  g^{\{u\}\{u\}}_< \, g^{\{v\}\{v\}}_>
\!\!\!\sum_{\scriptsize\begin{array}{c} \alpha\cup\bar\alpha=\{u\}\\ \gamma\cup\bar\gamma=\{v\}\end{array}}
(-1)^{P_\alpha+P_\gamma} d^{\alpha}a^{\bar\alpha}a^{\gamma}d^{\bar\gamma} h^{\alpha \gamma} h^{\bar\gamma \bar\alpha} h^{\alpha \bar\alpha} h^{\bar\gamma\gamma} \det t^{\alpha \gamma} \det t^{\bar\gamma \bar\alpha}  \, ,
\la{genscalar}
\eeq
where we are using the notation introduced in \eqref{morenotation} and $(-1)^{P_{\a}}$ is defined as a sign of the permutation of the ordered set $\{v\}$ which gives $\alpha\cup\bar\alpha$.  Note that the sum runs over all partitions $\a \cup \bar \a = \{u\}$ and $\gamma \cup \bar \gamma = \{v\}$, such that the number of elements in $\a$ and the number of elements in $\gamma$ are equal.  We can use the conversion factor in \eqref{conversion} to obtain the corresponding expression in the coordinate base as:
 \beqa
\,^\co\<\{v\}|\{u\}\>^\co &=& \frac{1}{d^{\{u\}} a^{\{v^*\}} g^{\{u+\frac{i}{2}\}} g^{\{v^*-\frac{i}{2}\}} f_<^{\{u\}\{u\}} \, f_>^{\{v^*\}\{v^*\}}} \, S^{\al}_N(\{v^*\},\{u\})
\la{innercoor}
\eeqa
where $S_N^{\al}$ is defined in (\ref{theguy}). Recall that $\mathcal{C}(u^*)=-\left[ \mathcal{B}(u)\right]^\dagger$ so that
\beq
S^{\al}_N(\{v^*\},\{u\})=(-1)^N\,^\al\<\{v\}|\{u\}\>^\al \,.
\eeq
\subsubsection*{New recursion relation in the coordinate basis}
Alternatively, we can define the general scalar product recursively, as in \eqref{newrecal}. Using the conversion factor \eqref{conversion}, the new recursion relation in the coordinate basis reads
\beqa
&&S^\co_N \(\{v_1,\dots,v_N\},\{u_1,\dots,u_N\}\) = \sum_{n} b_n^\co \, S^\co_{N-1} (\{v_1,\dots, \hat v_n,\dots,v_N\},\{\hat u_1, u_2,\dots,u_N\})\nn  \\
&&- \sum_ {n<m}  c_{n,m}^\co \, S^\co_{N-1}(\{ u_1,v_1,\dots \hat v_{n},  \dots , \hat v_{m},\dots v_N \}, \{\hat u_1,u_2,\dots,u_N\}) \, ,
\la{newrecco}
\eeqa
with the coefficients being in this case
\beqa
b^\co_n &= & \frac{\displaystyle \prod\limits_{j\neq n}^N f(u_1-v_j) \prod\limits_{j<n}^N S(v_j,v_n)  - \frac{a(u_1) d(v_n)}{d(u_1)a(v_n)}\prod\limits_{j\neq n}^N f(v_j-u_1) \prod\limits_{j>n}^N S(v_n,v_j)  }{\displaystyle \frac{g(u_1+\frac{i}{2}) \, g(v_n-\frac{i}{2}) }{g(u_1-v_n)}  \prod\limits_{j\neq1}^N f(u_1-u_j) }  \nn \\
c^\co_{n,m} &=&  \frac{\displaystyle S(v_{m},v_{n}) \frac{d(v_n)}{a(v_n)}\prod\limits_{j>n} S(v_{n},v_j) \prod\limits_{j<m} S(v_j,v_m)  + \frac{d(v_m)}{a(v_n)} \prod\limits_{j>m} S(v_{m},v_j) \prod\limits_{j<n} S(v_j,v_{n}) }{\displaystyle \[ \frac{d(u_1) \, g(u_1 + \frac{i}{2}) \,  \, g(v_{n}-\frac{i}{2}) \, g(v_{m}-\frac{i}{2}) \prod\limits_{j\neq 1} f(u_1-u_j) }{a(u_1) \, g(u_1-\frac{i}{2}) \, g(u_1-v_{n}) \, g(u_1-v_{m}) \prod\limits_{n\neq n,m} f(v_j - u_1) }\]}  \nn
\eeqa
where we used the usual $SU(2)$ S-matrix
$
S(u,v) = f(u-v)/{f(v-u)}
$.

\subsubsection*{Usual recursion relation}
Finally, for completeness, we write the recursion relation for the general scalar product known in the literature \cite{Korepin}
\begin{align} \la{oldrecal}
S^\al_N\[a(x),d(x)\]= \, &\sum_{n=1}^Ng(u_1-v_n) \, \prod_{j\neq 1}^N g(u_1 -u_j) \, \prod_{k\neq n}^N g(v_k-v_n)_{k \neq n} \\
&\times \left\{ a(v_n) d(u_1)S^\al_{N-1}\[a(x)h(u_1-x),d(x)h(x-v_n)\] \right. \nn\\
& \quad - \left. a(u_1)d(v_n)S^\al_{N-1}\[a(x)h(v_n-x),d(x)h(x-u_1)\] \right\}~,\nn
\end{align}
where the sets of magnons entering $S^\al_{N-1}$ are obtained from the set entering $S^\al_N$ by omitting $u_1$ and $v_n$. The difference between both sides of \eqref{oldrecal} lies in the action of the operators $\mathcal A(u)$ and $\mathcal D(u)$ on the vacuum.  For $S^\al_N$, this action is simply given by $a(u)$ and $d(u)$, see \eqref{ad}.  For the first $S^\al_{N-1}$ in \eqref{oldrecal}, this action is given by $a(u)h(u_1-u)$ and $d(u)h(u-v_n)$, while for the second $S^\al_{N-1}$ it is $a(u)h(v_n-u)$ and $d(u)h(u-u_1)$. That is, this recursion relation uses somehow generalized objects at each recursion stage, see also discussion at the end of section \ref{newsec}.

\subsubsection*{Scalar product with a Bethe eigenstate}
If the set of rapidities $\{u\}$ satisfies the Bethe equations \eqref{BAE}, while $\{v\}$ are arbitrary complex numbers, the scalar product simplifies to \cite{Slavnov}
\beq
S^\al_N (\{v\},\{u\}) =g^{\{u\}\{u\}}_> \, g^{\{v\}\{v\}}_< \, d^{\{u\}} \det_{j,k}\Omega(u_j,v_k)~,
\la{GenBethe}
\eeq
where
\beqq
\Omega(u_j,v_k)=a(v_k)t(u_j-v_k) h^{\{u-v_k\}}-(-1)^Nd(v_k)t(v_k-u_j) h^{\{v_k-u\}} \, .
\eeqq
In order to derive \eqref{GenBethe}, one starts with \eqref{genscalar} and makes use of the fact that the Bethe equations \eqref{BAE} for $\{u\}$ can be written as
\beqq
\frac{a(u_j)}{d(u_j)}=(-1)^{N-1} {h^{\{u_j-u\}} \over h^{\{u-u_j\}}}.
\eeqq
We refer the reader to \cite{Slavnov} for the details of the derivation.

\subsection{Scalar products in $SL(2)$ and $SU(1|1)$} \la{sl2su11}
We now give several formulas for the scalar products in the other rank-1 subsectors of $\N=4$ SYM.  All these formulas can be checked against the brute force computation in the coordinate basis using the corresponding wave function.  The latter is constructed in exactly the same way as in the $SU(2)$ case, see \eqref{waveN}, except that now the plave wave coefficients are obtained from products of the following S-matrices:\footnote{Recall that the $SL(2)$ spin chain is non-compact, meaning that there is no bound on the number of excitations that we can have at a given site in the spin chain. That is, a generic $SL(2)$ state is of the form
\beqq
|\Psi\>_{SL(2)} = \sum_{1 \le n_1\leq n_2\leq\dots \leq n_N\le L} \psi_{SL(2)}(n_1,\dots,n_N) |n_1,\dots , n_N\> \, .
\eeqq
It is important to take into account these new limits of summation when computing the scalar product.
}
\beqq
S_{SL(2)}(u_b,u_a) = \frac{u_b -u_a -i}{u_b -u_a +i} \, , \qquad \quad  S_{SU(1|1)} = -1 \,.
\eeqq

\subsection*{$SL(2)$}

The general inner product for $SL(2)$ spin chains obeys the same recursion relation as that for $SU(2)$ \eqref{newrecal}, but with the following expressions for the building blocks
\beqq
a(u) = \(u-\frac{i}{2}\)^L, \quad d(u)=\(u+\frac{i}{2}\)^L,
\eeqq
\beqq
f(u) = \frac{u+i}{u} \, , \quad g(u)=\frac{i}{u} \, , \quad h(u)=\frac{u+i}{i} \, , \quad t(u)=\frac{-1}{u(u+i)} \, .
\eeqq
However, the factor that relates the general inner product in the algebraic and coordinate bases is different than that for $SU(2)$.  Namely
\beq
\,^\co\<\{v\}|\{u\}\>^\co =  \frac{(-1)^{N}}{d^{\{v^*\}} a^{\{u\}} g^{\{u+\frac{i}{2}\}} g^{\{v^*-\frac{i}{2}\}} f^{\{u\}\{u\}}_> \, f^{\{v^*\}\{v^*\}}_<} \, S^{\al}_N(\{v^*\},\{u\}) \, .
\la{convsl2}
\eeq

Recall that the $SL(2)$ Bethe equations are
\beq
e^{i \phi_j}=1 \qquad \text{where} \qquad e^{i \phi_j}\equiv \(\frac{u_j+i/2}{u_j-i/2}\)^L  \prod_{k\neq j}^N \frac{u_j-u_k+i}{u_j-u_k-i} \,.
\la{BAEsl2}
\eeq
Then, if the set of rapidities $\{u\}$ satisfies these Bethe equations, the scalar product simplifies to
\beq
S^\al_N (\{v\},\{u\}) = g^{\{u\}\{u\}}_> \, g^{\{v\}\{v\}}_< \, d^{\{u\}}\det_{j,k}\Omega(u_j,v_k)~,
\eeq
where the matrix $\Omega$ is
\beqq
\Omega(u_j,v_k)=a(v_k)  t(u_j-v_k) h^{\{u-v_k\}}-(-1)^N d(v_k) t(v_k-u_j) h^{\{v_k-u\}} \, .
\eeqq
Finally, the norm in the coordinate basis is given by our conjectured formula \eqref{normconj}
\beq
\N_\co ={1 \over g^{\{u+{i\over 2}\}} g^{\{u-{i \over 2}\}}}\, \det_{j,k} \partial_{j} \phi_k \,,
\eeq
with $\phi_j$ defined in \eqref{BAEsl2}.

\subsection*{$SU(1|1)$}
The general inner product for $SU(1|1)$ spin chains can be computed using the sum over partitions formula of $SU(2)$ \eqref{genscalar}, but with the following expressions for the building blocks
\beqq
a(u)=\(u+\frac{i}{2}\)^L \, , \quad d(u)=\(u-\frac{i}{2}\)^L \, ,
\eeqq
\beqq
f(u) = \frac{1}{u}, \quad g(u)=\frac{i}{u}, \quad h(u)=\frac{1}{i}, \quad t(u)=-\frac{1}{u}
\eeqq
In this case, the relation between the scalar product in the algebraic and coordinate bases is exactly the same as that for $SU(2)$ \eqref{innercoor}. The $SU(1|1)$ Bethe equations are
\beq
e^{i \phi_j}=1 \qquad \text{where} \qquad e^{i \phi_j}\equiv \(\frac{u_j+i/2}{u_j-i/2}\)^L  \,.
\la{BAEsu11}
\eeq
Then, if the set of rapidities $\{u\}$ satisfies these Bethe equations, the scalar product simplifies to
\beq
S^\al_N (\{v\},\{u\}) = g^{\{u\}\{u\}}_> \, g^{\{v\}\{v\}}_< \, d^{\{u\}} \det\Omega(u_j,v_k)~,
\eeq
where the matrix $\Omega$ is now given by
\beqq
\Omega(u_j,v_k)=(-i)^{N} \[a(v_k)t(u_j-v_k) +d(v_k)t(v_k-u_j) \].
\eeqq
Finally, the norm in the coordinate basis is again given by formula \eqref{normconj} with $\phi_j$ defined in \eqref{BAEsu11}.

\section{The BMN limit} \la{BMNsec}
In this section we consider a simple limit
when the lengths of operators $L_a$ scale to infinity
whereas the numbers of magnons $M_a$ are fixed. We also assume that
the momenta of the magnons are small (or equivalantly the Bethe roots are large).
That is, similarly to the thermodynamical limit we have $u_k\simeq L_a$. These states
describe small quasi-classical fluctuations about the BMN point-like string. For related papers, see \cite{Beisert:2002bb,CorrelatorsinBMN}.

In this limit in the Bethe ansatz equations one can replace the $S$ matrix factor
by $1$ which leads to the trivial quantization condition:
\beq
u_k\simeq \frac{L}{2\pi n_k}
\eeq
where $n_k$ is an integer. In general one should assume that all $n_k$'s are different.
Otherwise there is a degeneracy which is lifted at the next order only.\footnote{ Namely,
for the $M_n$ roots with $n_k=n$ to the next order in $1/\sqrt L$ are \cite{Dhar}
\beq
u_l\simeq \frac{1}{2\pi n}\(L+i z_{l}\sqrt{2L}\)\;\;,\;\;l=1,\dots,M_n
\eeq
where $z_l$ are the zeros of the Hermite polynomial, ${\rm H}_{M_n}(z_k)=0$.
In what follows we assume that the roots are large and well separated between each other.}

As it was shown in the main text,
in order to construct all the structure constants one needs only tree key structures
 denoted by $\caA,\caB,\caC$. They can be easily expanded
in the near BMN limit\footnote{It is also very simple to derive them directly from their definition and the form of the wave function (\ref{waveN}). Recall that up to trivial factors $\caB, \caA, \caC$ are related to the norm of eigenstates, inner product with vacuum descendants and general inner product respectively. When the $S$-matrix is replaced by $1$ all these quantities are very simple to compute directly without any fancy integrability machinery.}
\beqa
{\caA}(l|\{ u\})&\simeq& \prod_{k=1}\(1-e^{\frac{l}{iu_k}}\)\\
{\caB}(\{u\})&\simeq& \sqrt{\(\bea{c} L-2M\\N-M\eea\)\frac{1}{L}}\prod^M_{k=1}\frac{i\sqrt{L}}{u_{k}}\\
{\caC}(l|\{ u\},\{ v\})&\simeq& \sum_{\sigma}
\prod_{k=1}\frac{e^{\frac{il}{u_k}}-e^{\frac{il}{v_{\sigma_k}}}}{i(u_k-v_{\sigma_k})}\;.
\eeqa
Having these quantities at hand we simply combine them into structure constants using
the expressions in table \ref{finalresults}. For example,
\beq
\; C_{123}^{\bullet\bullet\circ}\simeq
\frac{
\prod^{M_1}_{k=1}\frac{v_{k}}{i\sqrt{L_1}}
\prod^{M_2}_{k=1}\frac{u_{k}}{i\sqrt{L_2}}
}{
\sqrt{\frac{{L_1-2M_1 \choose N_1-M_1}{L_2-2M_2 \choose N_2-M_2} {L_3 \choose N_3}}{L_1L_2L_3}}}
\sum\limits_{\scriptsize\begin{array}{c}\alpha\cup\bar\alpha=\{u\}\\|\bar \alpha|={l_{13}}\end{array}}
\!\!\!\!\sum_{\sigma}
\[\prod_{k=1}^{|\alpha|}\frac{e^{\frac{il_{12}}{\alpha_k}}-e^{\frac{il_{12}}{v_{\sigma_k}}}}{i(\alpha_k-v_{\sigma_k})}\]
\[\prod_{k=1}^{|\bar \alpha|}{e^{\frac{iL_1}{\bar \alpha_k}}-e^{\frac{il_{12}}{\bar \alpha_k}}}\]\;.
\eeq

\section{Circular string details}\la{AppQC}
The density of roots in the scaling limit and the resolvent are known explicitly \cite{KMMZ}\footnote{The algebraic curve for the general $SU(2)$ circular string at strong coupling was similarly written down in \cite{GV1}.}
\beqa
\rho(u)&=&\frac{i}{2\pi u}\sqrt{(2\pi n u-Le^{+i\phi})(2\pi n u-Le^{-i\phi})}\;,\\
G(u)&=&\int \frac{\rho(v)}{u-v}dv=
\frac{1}{2 u}\sqrt{(2\pi n u-Le^{+i\phi})(2\pi n u-Le^{-i\phi})}+\frac{L}{2u}-\pi n\;
\eeqa
where the filling fraction $M_2/L_2\equiv\alpha$ is related to $\phi$ by  $\alpha=\sin^2\frac\phi2$.
Substituting these expressions into \eq{chapeau} we get
\beqa
&&\frac{1}{L_2}\log{{{\caA}(l|\{v\})
\over {\caB}_2}}\simeq\frac{1}{2} \alpha  \log \left(\frac{e \beta ^2 \sin ^2(r)}{\alpha  r^2}\right)
+\alpha ^2 \left(\frac{r^2 \left(3 (\beta -1) \beta  \csc ^2(r)+1\right)}{6 \beta ^2}-\frac{1}{4}\right)+\\
\nn&&\alpha ^3 \left(\frac{r^4 \left(1-15 (\beta -1)^2 \beta ^2 (\cos (2 r)+2) \csc ^4(r)\right)}{90 \beta
   ^4}+\frac{(\beta -1) (2 \beta -1) r^3 \cot (r) \csc ^2(r)}{3 \beta ^2}-\frac{1}{12}\right)\\
   \nn &&-\frac{\alpha^4}{24}-
   \alpha ^4 \frac{(\beta -1)^2 (2 \beta -1) r^5 (11 \cos (r)+\cos (3 r)) }{12 \sin ^5(r)\beta
   ^3}\\
\nn&&+\alpha ^4 \frac{r^6 \left(63 (\beta -1)^3 \beta ^3 (26 \cos (2 r)+\cos (4 r)+33) \csc ^6(r)+8\right)}{4536
   \beta ^6}\\
\nn&&+\alpha ^4\frac{r^4 \left(120 (\beta -1) \beta  (5 (\beta -1) \beta +1) (\cos (2 r)+2) \csc
   ^4(r)-8\right)}{1440 \beta ^4}
\eeqa
where $\beta=l_{12}/L_2$ and $r=\pi n\beta$.

\section{Mathematica codes}
In this appendix we provide some \verb"Mathematica" codes for computing the structure constants described in the main text. We implement both the brute force computation (\ref{3ptbf}) as well as the final analytic results of table \ref{finalresults}. Some examples are presented at the end of this appendix.

\subsection*{Structure constants by brute force (\ref{3ptbf})}
{\small
\verb"Off[Det::matsq];   Le = Length;"\\
\verb"f[u_] = 1 + I/u;   g[u_] = I/u;   n[0] = 0;"\\
\verb"S[x_, y_] := (x - y + I)/(x - y - I)"\\
\verb"Wave[l_List] := Block[{p = Permutations[Range[Le[l]]], i, j}, Sum[A[p[[i]]]"\\
\verb" Product[((l[[p[[i, j]]]] + I/2)/(l[[p[[i, j]]]] - I/2))^n[j], {j, 1, Le[l]}],"\\
\verb" {i, 1, Le[p]}] //. {A[{a___, b_, c_, d___}] :> S[l[[b]], l[[c]]] A[{a, c, b, d}]"\\
\verb" /; b > c} /. {A[a___] :> 1 /; a == Range[Le[a]]}];"\\
\verb""\\
\verb"normbf[L_, l_List] := Sum[(Wave[l] /. Complex[a_, b_] -> a - I b) Wave[l],"\\
\verb"Evaluate[Sequence @@ Table[{n[j], n[j - 1] + 1, L}, {j, Le[l]}]]]"\\
\verb""\\
\verb"dphi[L_,l_List] := Det@Table[-If[i==j,L/(l[[i]]^2+1/4)-Sum[2/((l[[i]]-l[[k]])^2+1),"\\
\verb" {k,Le[l]}],0]-2/(1+(l[[i]]-l[[j]])^2),{i,Le[l]},{j,Le[l]}] /. Det[{}] -> 1"\\
\verb"prefactor[l_List] := Product[1/(g[l[[j]] + I/2]g[l[[j]] - I/2]), {j, 1, Le[l]}]*"\\
\verb" Product[f[l[[j]] - l[[i]]]/f[Conjugate[l[[j]] - l[[i]]]], {i, 1, Le[l]},"\\
\verb" {j, i + 1, Le[l]}]"\\
\verb"normdet[L_, l_List] := prefactor[l] dphi[L, l]"\\
\verb""\\
\verb"C123[L1_, N1_, L2_, N2_, L3_, N3_, l1_List, l2_List, l3_List] :="\\
\verb" Block[{i, j, psis, norms, limits},"\\
\verb" psis = (Wave[l1] /. n[j_] :> L1 - N3 + j - N2 /; j > N2)"\\
\verb" (Wave[l2] /. n[j_] -> L2 + 1 - n[N2 - j + 1]) (Wave[l3] /. n[j_] -> j);"\\
\verb" norms = normdet[L1, l1] normdet[L2, l2] normdet[L3, l3];"\\
\verb" limits = Sequence @@ Table[{n[j], n[j - 1] + 1, L1 - N3}, {j, N2}];"\\
\verb" Sqrt[L1 L2 L3/norms] If[limits === Sequence[], psis, Sum[psis, limits]]]"\\
\verb""\\
}
Note that the norms can be computed by brute force using {\verb"normbf"}, as in \eqref{normbf}, or using {\verb"normdet"}, which is just the implementation of \eqref{normcoord}. We choose to use the latter as it is computationally much more efficient.

\subsection*{Structure constants from table \ref{finalresults}}
{\small
\verb"Off[Det::matsq];   Le = Length;"\\
\verb"f[u_] = 1 + I/u;    g[u_] = I/u;    h[u_] = f[u]/g[u];    t[u_] = g[u]^2/f[u];"\\
\verb"f[l1_List, l2_List] := Product[f[l1[[j1]] - l2[[j2]]], {j1, Le[l1]}, {j2, Le[l2]}]"\\
\verb"h[l1_List, l2_List] := Product[h[l1[[j1]] - l2[[j2]]], {j1, Le[l1]}, {j2, Le[l2]}]"\\
\verb"fs[l1_List] := Product[f[l1[[j1]] - l1[[j2]]], {j1, Le[l1]}, {j2, j1 + 1, Le[l1]}]"\\
\verb"gs[l1_List] := Product[g[l1[[j1]] - l1[[j2]]], {j1, Le[l1]}, {j2, j1 + 1, Le[l1]}]"\\
\verb"fb[l1_List] := Product[f[l1[[j1]] - l1[[j2]]], {j1, Le[l1]}, {j2, j1 - 1}]"\\
\verb"gb[l1_List] := Product[g[l1[[j1]] - l1[[j2]]], {j1, Le[l1]}, {j2, j1 - 1}]"\\
\verb"gp[l_List] := Times @@ g[l + I/2];    gm[l_List] := Times @@ g[l - I/2]"\\
\verb"a[l_List] := Times @@ ((l + I/2)/(l - I/2))"\\
\verb"sign[a_List, ab_List, v_List] := Signature[Join[a, ab]] Signature[v]"\\
\verb"Dvd[ls_List] := ({Complement[ls, #1], #1} & ) /@ Subsets[ls, {0, Le[ls]}]; "\\
\verb"dphi[L_,l_List]:=Det@Table[-If[i==j,L/(l[[i]]^2+1/4)-Sum[2/((l[[i]]-l[[k]])^2+1),"\\
\verb" {k,Le[l]}],0]-2/(1+(l[[i]]-l[[j]])^2),{i,Le[l]},{j,Le[l]}] /. Det[{}] -> 1"\\
\verb"dett[l1_List,l2_List]:=Det@Table[t[l1[[i]]-l2[[j]]],{i,Le[l1]},"\\
\verb" {j,Le[l2]}] /. Det[{}] -> 1"\\
\verb"A[L_, ls_List] := Block[{dv = Dvd[ls],al,alb}, Sum[al = dv[[i, 1]];"\\
\verb" alb = dv[[i, 2]]; (-1)^Le[al] f[al, alb]/a[al]^L, {i, Le[dv]}]];"\\
\verb"B[L_,N_,l_List]:=gm[l]fs[l]Sqrt[fb[l]dphi[L,l]Binomial[L-2Le[l],N-Le[l]]"\\
\verb" /(fb[Conjugate[l]]L)]/Sqrt[gp[l]gm[l]]"\\
\verb"T[n_,u_List,v_List]:=Block[{dv=Dvd[v],du=Dvd[u],al,alb,be,beb},gs[u]gb[v]Sum["\\
\verb" al=dv[[i,1]];alb=dv[[i,2]];be=du[[j,1]];beb=du[[j,2]];If[Le[al]==Le[be]&&"\\
\verb" Le[alb]==Le[beb],sign[al,alb,v]sign[be,beb,u]dett[be,al]dett[alb,beb]a[beb]^n"\\
\verb" a[al]^n h[be,al]h[alb,beb]h[be,beb]h[alb,al],0],{i,Le[dv]},{j,Le[du]}]]"\\
\verb"" \\
\verb"Cooo[L1_,N1_,L2_,N2_,L3_,N3_]:=(Binomial[L1-N1+N2,N2]/"\\
\verb" (B[L1,N1,{}]B[L2,N2,{}]B[L3,N3,{}]))"\\
\verb"Cxoo[L1_,N1_,L2_,N2_,L3_,N3_,l1_List]:=(Binomial[L1-N1+N2-Le[l1],N2]"\\
\verb" A[N1-N2,l1]/(B[L1,N1,l1]B[L2,N2,{}]B[L3,N3,{}]))"\\
\verb"Coxo[L1_,N1_,L2_,N2_,L3_,N3_,l2_List]:=(Binomial[L1-N1+N2-Le[l2],N2-Le[l2]]"\\
\verb" A[L1-N1+N2,l2]/(B[L1,N1,{}]B[L2,N2,l2]B[L3,N3,{}]))"\\
\verb"Coox[L1_,N1_,L2_,N2_,L3_,N3_,l3_List]:=(Binomial[L1-N1+N2,N2]"\\
\verb" A[N1-N2,l3]/(B[L1,N1,{}]B[L2,N2,{}]B[L3,N3,l3]))"\\
\verb"Coxx[L1_,N1_,L2_,N2_,L3_,N3_,l2_List,l3_List]:=(Binomial[L1-N1+N2-Le[l2],N2-Le[l2]]"\\
\verb" A[L1-N1+N2,l2]A[N1-N2,l3]/(B[L1,N1,{}]B[L2,N2,l2]B[L3,N3,l3]))"\\
\verb"Cxox[L1_,N1_,L2_,N2_,L3_,N3_,l1_List,l3_List]:=(Binomial[L1-N1+N2-Le[l1],N2]"\\
\verb" A[N1-N2,l1]A[N1-N2,l3]/(B[L1,N1,l1]B[L2,N2,{}]B[L3,N3,l3]))"\\
\verb"Cxxo[L1_,N1_,L2_,N2_,L3_,N3_,l1_List,l2_List]:=(dv=Dvd[l1];"\\
\verb" Sum[al=dv[[i,1]];alb=dv[[i,2]];If[Le[al]==N1-N2,a[al]^L1 f[alb,al]A[N1-N2,al]"\\
\verb" T[L1-N1+N2,alb,l2],0],{i,Le[dv]}]/(B[L1,N1,l1]B[L2,N2,l2]B[L3,N3,{}]))"\\
\verb"Cxxx[L1_,N1_,L2_,N2_,L3_,N3_,l1_List,l2_List,l3_List]:=(A[N1-N2,l3]B[L3,N3,{}]"\\
\verb" Cxxo[L1,N1,L2,N2,L3,N3,l1,l2]/B[L3, N3, l3])"
}

\subsection*{Examples}
Let us show how to use the code above to compute the most general case $C^{\bullet \bullet \bullet}_{123}$ in table \ref{finalresults} for some three-point function configuration.  In this example, the sets of rapidities \verb"us", \verb"vs", \verb"ws" satisfy the Bethe equations of operators $\O_1$, $\O_2$ and $\O_3$, respectively, with the lengths and number of excitations indicated in each case. Evaluating \\ \\
{\small
\verb"L1=9; N1=4; L2=9; N2=2; L3=4; N3=2;"\\
\verb"us = {0.414080361016 - 0.993048580811 I, 0.409292874229,"\\
\verb"      0.414080361016 + 0.993048580811 I, -0.131911999475};"\\
\verb"vs = {-0.207106781187, 0.207106781187};"\\
\verb"ws = {-0.288675134595, 0.288675134595};"\\
\verb"C123[L1,N1,L2,N2,L3,N3,us,vs,ws]"\\
\verb"Cxxx[L1,N1,L2,N2,L3,N3,us,vs,ws]"\\
}\\
we obtain a perfect match between the brute force computation (\ref{3ptbf}) and the analytic prediction of table \ref{finalresults}. For these values of the Bethe roots we find
\beqq
C^{\bullet \bullet \bullet}_{123} = -0.473631 + 0.079146 \, i \, .
\eeqq

We can also use the codes above to get analytic results. For example,  to reproduce $C_{123}^{\circ \bullet \circ}$ for the case of two magnons $N_2=2$, see equation \eqref{Twomag}, we would simply run
{\small
\verb""\\
\verb""\\
\verb"ClearAll[L1,N1,L2,N2,L3,N3]"\\
\verb"fsi=FullSimplify[#,{0<p<\[Pi]/2}]&;"\\
\verb"fsi2=#//.{Exp[a_]-Exp[b_]:>Exp[(a + b)/2]2 Sinh[(a-b)/2//fsi],"\\
\verb" Exp[a_]-1:>Exp[a/2]2 Sinh[a/2//fsi],Exp[a_]:>Exp[a//fsi]}&;"\\
\verb"fsi@Coxo[L1,N3+2,L2,2,L3,N3,{u,-u}]/.u->Cot[p/2]/2//TrigToExp//Factor//fsi//fsi2"\\
}

\section{Data} \la{dataAp}
In this appendix we will produce one-loop data for the future. Two things need to be done: we need to contract the external wave functions of the three operators using Hamiltonian insertions \cite{Okuyama:2004bd,Roiban:2004va,Alday:2005nd} and we need to take into account the $\mathcal{O}(\lambda)$ correction to the wave function, which is a two-loop effect. The latter effect was not taken into account in previous works \cite{Okuyama:2004bd,Roiban:2004va,Alday:2005nd} but its importance was mentioned in the conclusions of \cite{Okuyama:2004bd}.

For simplicity we shall consider two kind of operators only: BPS operators and operators with two impurities with opposite momentum
\beq
u=\pm\( \frac{1}{2}\cot\frac{\pi n}{L-1} + \frac{\lambda}{8\pi^2} \frac{L }{L-1}\,\sin \frac{2\pi n}{L-1}  +\mathcal{O}(\lambda^2)\)
\eeq
and energy
\beq
\Delta= \frac{\lambda}{\pi^2} \sin^2 \frac{n\pi}{L-1}-\frac{\lambda^2}{4\pi^4} \frac{1}{L-1}\(1+L+2\cos \frac{2\pi n}{L-1}\) \sin^4\frac{\pi n}{L-1}+ \mathcal{O}(\lambda^3)  \,.
\eeq
The integer $n=1,\dots,L/2-1$ is called the mode number. These states are denoted by $(n,L)$ and they diagonalize the two-loop Hamiltonian of \cite{BKS}. We take\footnote{In this section we shall not keep track of the overall phase of the structure constant but only its absolute value.} ($g^2=\lambda/16\pi^2$)
\begin{align*}
(1,4)&= \Tr \[Z,X\]^2  \, , \\
(1,5)&= \Tr \(Z^3X^2\)-\Tr \(Z^2XZX\) \, , \\
(1,6)&=  \Tr \(Z^4X^2\)- \frac{3-\sqrt{5}}{2}(1+g^2)\, \Tr \(Z^3XZX\)+\frac{1-\sqrt{5}+(3-\sqrt{5})g^2}{2}\,\Tr\(Z^2XZ^2X\)  \, ,
\\
(2,6)&=   \Tr \(Z^4X^2\)- \frac{3+\sqrt{5}}{2}(1+g^2)\, \Tr \(Z^3XZX\)+\frac{1+\sqrt{5}+(3+\sqrt{5})g^2}{2}\,\Tr\(Z^2XZ^2X\) \,
\\
(1,7)&= \Tr \(Z^5X^2\) -\frac{g^2}{2} \, \Tr \(Z^4XZX\)  -\frac{1}{2} (2-g^2)\, \Tr \(Z^3XZ^2X\)  \, ,    \\
(2,7)&= \Tr \(Z^5X^2\) - \(2+\frac{3g^2}{2}\) \, \Tr \(Z^4XZX\) + \(1+ \frac{3g^2}{2}\) \, \Tr \(Z^3XZ^2X\)  \, ,    \\
(1,8)&= \Tr \(Z^6X^2\) +  \[ \sec \(\frac{\pi}{7}\) \sin \(\frac{\pi}{14}\) - 0.489997 \, g^2 \] \, \Tr \(Z^5XZX\) \\
& \quad - \, \[ \sec \(\frac{\pi}{7}\) \sin \(\frac{3\pi}{14}\) - 0.206966 \, g^2 \] \, \Tr \(Z^4XZ^2X\) \\
&\quad - \[ \frac{1}{2} \sec \(\frac{\pi}{7}\) - 0.283031 \, g^2\] \, \Tr \(Z^3XZ^3X\) \, , \\
(2,8)&= \Tr \(Z^6X^2\) - \[ \cos \(\frac{\pi}{7}\) \csc \(\frac{3\pi}{14}\) +1.117057 \, g^2 \] \, \Tr \(Z^5XZX\) \\
& \quad - \, \[ \sin \(\frac{\pi}{14}\) \csc \(\frac{3\pi}{14}\) - 1.023185 \, g^2\] \, \Tr \(Z^4XZ^2X\)  \\
& \quad +  \, \[\frac{1}{2} \csc \(\frac{3\pi}{14}\) + 0.093872 \, g^2\] \, \Tr \(Z^3XZ^3X\) \, ,  \\
(3,8)&= \Tr \(Z^6X^2\) - \[\csc \(\frac{\pi}{14}\) \sin \(\frac{3\pi}{14}\)  + 3.392947 \,g^2\] \, \Tr \(Z^5XZX\) \\
& \quad + \[\csc \(\frac{\pi}{14}\) \cos \(\frac{\pi}{7}\) + 8.769849 \,g^2\] \, \Tr \(Z^4XZ^2X\) \\
&\quad - \[ \frac{1}{2} \csc \(\frac{\pi}{14}\) + 5.376903 \, g^2 \] \, \Tr \(Z^3XZ^3X\) \, .
\end{align*}

\begin{table}
\begin{center}
\footnotesize
\begin{tabular}{ | c c c | c || c c c | c |  }
  \hline
  $\O_1$ & $\O_2$ & $\O_3$ & $r$ & $\O_1$ & $\O_2$ & $\O_3$ & $r$ \\ \hline \hline

\color{blue}   (1,4) &\color{red} \bps &\color{red}  \bps & $-6$ &\color{blue}   (3,8) & \color{red} \bps &\color{red}  \bps & $-9.561263$ \\
 \color{red}  \bps &\color{red}  \bps &\color{blue}   (1,4) & $-6$ &  $\color{red} [2,4]$ & \color{blue}  (3,8) & \color{red} $[0,4]$ & \color{darkgreen}$-23.396074$ \\
\color{blue}    (1,5) & \color{red} \bps &\color{red}  \bps & $-4$ & \color{red}  $[4,4]$ &\color{blue}   (3,8) & \color{red} $[2,8]$ & $-9.561263$  \\
\color{red}   \bps &\color{red}  \bps &\color{blue}   (1,5) & $-4$ &\color{red}   $[4,8]$ &\color{blue}   (3,8) &\color{red}  $[2,4]$ & $-9.561263$  \\

  \color{blue} (1,6) & \color{red}  \bps &\color{red}   \bps & $-11/2 +13\sqrt{5}/10$ & \color{red}   $[3,5]$ &\color{blue}  (3,8) & \color{red}  $[1,5]$ & $-8.188323$ \\
  \color{red}  $[3,4]$ & \color{blue} (1,6) &\color{red}   $[1,4]$ & $3\sqrt{5}/10-7/2$ & \color{red}   $[4,6]$ &\color{blue}  (3,8) & \color{red}  $[2,6]$ & $-8.188323$  \\
 \color{red}   $[4,5]$ &\color{blue}  (1,6) &\color{red}   $[2,5]$ & $3\sqrt{5}/10-7/2$ &  \color{red}  $[3,6]$ & \color{blue} (3,8) & \color{red}  $[1,4]$ & $-5.330327$ \\
 \color{red}   \bps & \color{red}  \bps & \color{blue} (1,6) & $-11/2 +13\sqrt{5}/10$ &\color{red}   $[4,7]$ &\color{blue}  (3,8) &\color{red}   $[2,5]$ & $-5.330327$  \\

 \color{blue}   (2,6) & \color{red}  \bps & \color{red}  \bps & $-11/2 -13\sqrt{5}/10$ &\color{red}   $[4,5]$ &\color{blue}  (3,8) &\color{red}   $[2,7]$ & $-5.330327$  \\
 \color{red}    $[3,4]$ &\color{blue}  (2,6) & \color{red}  $[1,4]$ & $-3\sqrt{5}/10-7/2$ &\color{red}   $[3,4]$ &\color{blue}  (3,8) &\color{red}   $[1,6]$ & $-5.330327$  \\
\color{red}    $[4,5]$ &\color{blue}  (2,6) &\color{red}   $[2,5]$ & $-3\sqrt{5}/10-7/2$ &\color{red}   \bps &\color{red}   \bps &\color{blue}  (3,8) & $-9.561263$  \\
\color{red}     \bps &\color{red}   \bps &\color{blue}  (2,6) & $-11/2 -13\sqrt{5}/10$ &\color{red}   \bps & \color{blue} (1,4) & \color{blue} (1,4) & $-12$  \\

 \color{blue}    (1,7) &\color{red}   \bps &\color{red}   \bps & $-7/4$ &\color{red}   \color{red}  $[4,5]$ &\color{blue}  (1,5) & \color{blue} (1,4) & $-10$  \\
  \color{red}   $[3,4]$ & \color{blue} (1,7) & \color{red}  $[1,5]$ & $-2$ &\color{blue}  (1,4) &\color{red}   $[0,5]$ &\color{blue}  (1,5) & $-10$  \\
  \color{red}  $[3,5]$ &\color{blue}  (1,7) &\color{red}   $[1,4]$ & $-2$ &\color{red}   $[4,6]$ & \color{blue} (1,6) & \color{blue} (1,4) & $13\sqrt{5}/10-23/2$ \\
 \color{red}    $[4,5]$ &\color{blue}  (1,7) &\color{red}   $[2,6]$ & $-2$ &\color{blue}  (1,4) & \color{red}  $[0,6]$ &\color{blue}  (1,6) & $13\sqrt{5}/10-23/2$  \\
\color{red}    $[4,6]$ &\color{blue}  (1,7) &\color{red}   $[2,5]$ & $-2$ &\color{red}   $[4,7]$ & \color{blue} (1,7) & \color{blue} (1,4) & $-31/4$   \\
 \color{red}    \bps & \color{red}  \bps & \color{blue} (1,7) & $-7/4$ &\color{blue}  (1,4) &\color{red}   $[0,7]$ & \color{blue} (1,7) & $-31/4$  \\

 \color{blue}  (2,7) &\color{red}   \bps &\color{red}   \bps & $-27/4$ & \color{red}  $[4,8]$ & \color{blue} (1,8) &\color{blue}  (1,4) & $-7.237755$  \\
 \color{red}    \bps &\color{red}   \bps &\color{blue}  (2,7) & $-27/4$ &\color{blue}  (1,4) &\color{red}   $[0,8]$ &\color{blue}  (1,8) & $-7.237755$ \\

\color{blue}    (1,8) & \color{red}  $[0,8]$ &\color{red}   $[2,4]$ & $-1.237755$ &\color{blue}  (1,4) &\color{blue}  (1,8) &\color{red}   $[0,4]$ & \color{darkgreen}$5.123727$  \\
 \color{red}     $[2,4]$ &\color{blue}  (1,8) &\color{red}   $[{0,4}]$ & \color{darkgreen}$-4.525727$ & \color{red}   $[4,6]$ &\color{blue}  (2,6) &\color{blue}  (1,4) & $-13\sqrt{5}/10-23/2$ \\
 \color{red}     $[4,6]$ &\color{blue}  (1,8) &\color{red}   $[{2,6}]$ & $-1.513645$ &\color{blue}   (1,4) & \color{red}  $[{0,6}]$ &\color{blue}  (2,6) & $-13\sqrt{5}/10-23/2$ \\
 \color{red}     $[3,5]$ &\color{blue}  (1,8) &\color{red}   $[{1,5}]$ & $-1.513645$ & \color{red}   $[{4,7}]$ &\color{blue}  (2,7) &\color{blue}  (1,4) & $-51/4$ \\
\color{red}      $[3,4]$ &\color{blue}  (1,8) &\color{red}   $[{1,6}]$ & $-1.455824$ &\color{blue}   (1,4) &\color{red}   $[{0,7}]$ &\color{blue}  (2,7) & $-51/4$ \\
 \color{red}     $[3,6]$ &\color{blue}  (1,8) &\color{red}   $[{1,4}]$ & $-1.455824$ & \color{red}   $[{4,8}]$ &\color{blue}  (2,8) &\color{blue}  (1,4) & $-11.200983$ \\
 \color{red}     $[4,5]$ &\color{blue}  (1,8) &\color{red}   $[{2,7}]$ & $-1.455824$ & \color{blue}  (1,4) & \color{red}  $[{0,8}]$ &\color{blue}  (2,8) & $-11.200983$ \\
 \color{red}     $[4,7]$ &\color{blue}  (1,8) &\color{red}   $[{2,5}]$ & $-1.455824$ & \color{blue}  (1,4) &\color{blue}  (2,8) &\color{red}   $[{0,4}]$ & \color{darkgreen}$-1.554117$ \\
 \color{red}     $[4,4]$ &\color{blue}  (1,8) &\color{red}   $[{2,8}]$ & $-1.237755$ & \color{red}   $[{4,8}]$ &\color{blue}  (3,8) &\color{blue}  (1,4) & $-15.561263$ \\
 \color{red}     $[4,8]$ &\color{blue}  (1,8) &\color{red}   $[{2,4}]$ & $-1.237755$ &  \color{blue} (1,4) & \color{red}  $[{0,8}]$ &\color{blue}  (3,8) & $-15.561263$ \\
  \color{red}    \bps & \color{red}  \bps &\color{blue}  (1,8) & $-1.237755$ &\color{blue}   (1,4) & \color{blue} (3,8) &\color{red}   $[{0,4}]$ & \color{darkgreen}$-9.293748$ \\

   \color{blue}  (2,8) & \color{red}  $[{0,8}]$ &\color{red}   $[{2,4}]$ & $-5.200983$ &  \color{red}  $[{4,4}]$ &\color{blue}  (1,5) & \color{blue} (1,5) & $-8$ \\
    \color{red}  $[{2,4}]$ &\color{blue}  (2,8) &\color{red}   $[{0,4}]$ & \color{darkgreen}$-10.078199$ & \color{blue}  (1,5) & \color{red}  $[{0,6}]$ & \color{blue} (1,5) & $-8$ \\
   \color{red}   $[{4,7}]$ &\color{blue}  (2,8) &\color{red}   $[{2,5}]$ & $-7.213849$ &  \color{red}  $[{4,5}]$ &\color{blue}  (1,6) &\color{blue}  (1,5) & $3\sqrt{5}/10-15/2$ \\
   \color{red}   $[{4,5}]$ &\color{blue}  (2,8) &\color{red}   $[{2,7}]$ & $-7.213849$ & \color{blue}  (1,5) & \color{red}  $[{0,7}]$ &\color{blue}  (1,6) & $13\sqrt{5}/10-19/2$ \\
   \color{red}   $[{3,6}]$ &\color{blue}  (2,8) &\color{red}   $[{1,4}]$ & $-7.213849$ &  \color{red}  $[{4,6}]$ &\color{blue}  (1,7) &\color{blue}  (1,5) & $-6$ \\
   \color{red}   $[{3,4}]$ &\color{blue}  (2,8) &\color{red}   $[{1,6}]$ & $-7.213849$ & \color{blue}  (1,5) & \color{red}  $[{0,8}]$ &\color{blue}  (1,7) & $-23/4$ \\
   \color{red}   $[{4,8}]$ &\color{blue}  (2,8) &\color{red}   $[{2,4}]$ & $-5.200983$ &   \color{red}  $[{4,7}]$ &\color{blue}  (1,8) &\color{blue}  (1,5) & $-5.455824$ \\
   \color{red}   $[{4,4}]$ &\color{blue}  (2,8) &\color{red}   $[{2,8}]$ & $-5.200983$ &  \color{red}  $[{4,5}]$ &\color{blue}  (2,6) &\color{blue}  (1,5) & $-3\sqrt{5}/10-15/2$ \\
   \color{red}   $[{3,5}]$ &\color{blue}  (2,8) &\color{red}   $[{1,5}]$ & $-0.298032$ & \color{blue}  (1,5) & \color{red}  $[{0,7}]$ &\color{blue}  (2,6) & $-13\sqrt{5}/10-19/2$ \\
    \color{red}  $[{4,6}]$ &\color{blue}  (2,8) & \color{red}  $[{2,6}]$ & $-0.298032$ & \color{blue}  (1,5) & \color{red}  $[{0,8}]$ &\color{blue}  (2,7) & $-43/3$ \\
   \color{red}   \bps &\color{red}   \bps &\color{blue}  (2,8) & $-5.200983$ & \color{red}   \color{red}  $[{4,7}]$ &\color{blue}  (2,8) &\color{blue}  (1,5) & $-11.213849$ \\

  \hline
\end{tabular}
\caption{\small Data for the ratio between the one-loop and tree-level structure constants parametrized by $r$ in $N_c C_{123}=c_{ijk}^{(0)}\(1+g^2 r+\mathcal{O}(g^4)\)$. The notation $(n,L)$ indicates a two magnon operator with mode number $n$ and length $L$. The notation $[N,L]$ indicates a vacuum descendant with $N$ spin flips and total length $L$. The vacuum and excitation choice for each of the three operators is given in (\ref{Vacchoice}). When we write $\bps$ we can replace it by any $[N,L]$ with $N\le 4$ and $L \le 8$ (such that the three-point function exists); in this case the result is independent of $N$ and $L$. To make the table easier to read we colored the BPS operators in red and the non-BPS operators in blue. We also colored in green the ratios for configurations with $l_{13}=0$. For these cases (only for these cases) the mixing of $\O_2$ with double trace operators should in principle be included \cite{Beisert:2002bb,revieC} (we did not take this effect into account).}
\la{tabledata}
\end{center}
\end{table}

Note that for convenience, we gave the numerical values of exact expressions in the states shown above.  For example, in the last term of $\mathcal{O}_{3,8}$, we have
\beqq
5.376903 = \sqrt{\frac{9-8 \cos \(\frac{\pi}{7}\) +8 \sin \(\frac{3\pi}{14}\) }{44-54 \cos \(\frac{\pi}{7}\) + 78 \sin \(\frac{\pi}{14}\) -20 \sin \(\frac{3\pi}{14}\)} } \, .
\eeqq
We can also write any protected state with $N$ excitations and length $L$, which we denote by $[N,L]$, as
\beqq
[N,L] = \sum_{1 \leq  n_1 < n_2 < \dots < n_N \leq L} \Tr ( Z \dots \underset{\underset{n_1}{\downarrow}}{X} \dots \underset{\underset{n_2}{\downarrow}}{X}   \underset{\underset{\dots}{}}{\dots} \dots  \underset{\underset{n_N}{\downarrow}}{X} \dots Z)
\eeqq
For example:
\begin{align*}
[2,5] &= 5\, \Tr \(Z^3X^2\) + 5\, \Tr \(Z^2XZX\) \, , \\
[3,6] &= 6\, \Tr \(Z^3X^3\) + 6\, \Tr \(Z^2 X Z X^2\) + 6\, \Tr \(Z^2 X^2 Z X\) + 2\, \Tr \(ZXZXZX\) \, .
\end{align*}

All the states written in this appendix consider the vacuum as being $Z$ fields and the excitations as being the $X$ fields. In other words, they are good states for the operator $\mathcal{O}_1$. For the operators $\mathcal{O}_2$ and $\mathcal{O}_3$ we use the same states but with different scalars playing the role of vacuum and excitations, see (\ref{Vacchoice}).

Having computed the two-loop eigenstates $(n,L)$ and knowing how to write any BPS operators with a given length and number of excitations, we provide a list of ratios of one-loop structure constants to tree-level structure constants in table \ref{tabledata} involving these two kind of operators.

\end{document}